\documentclass[preprint2, letter]{proto}
\input{psfig}
\usepackage{times}
\newcommand{\refs}{\par\noindent\hangindent=1pc\hangafter=1}
\begin{document}

\title{\textbf{\LARGE THE DYNAMICAL STRUCTURE OF THE KUIPER BELT AND
ITS PRIMORDIAL ORIGIN}}

\author{\textbf{\large Alessandro  Morbidelli}}
\affil{\small\em Observatoire de la C\^ote d'Azur}

\author {\textbf{\large Harold F. Levison}}
\affil{\small\em Southwest Research Institute}

\author{\textbf{\large Rodney Gomes}}
\affil{\small\em Observat\'orio Nacional/MCT - Brazil}

\begin{abstract}
  \baselineskip = 11pt 
  \leftskip = 0.65in 
  \rightskip = 0.65in
  \parindent=1pc 
 
   {\small This chapter discusses the dynamical properties of the Kuiper belt
 population. Then, it focuses on the characteristics of the Kuiper
 belt that cannot be explained by its evolution in the framework of
 the current solar system. We review models of primordial solar
 system evolution that have been proposed to reproduce the Kuiper belt
 features, outlining advantages and problems of each of them.  \\~\\~\\~}
 
\end{abstract}

\section{Introduction}
\label{sec_intro}

{Since its discovery in 1992, the Kuiper belt has slowly revealed a
  stunningly complex dynamical structure.  This structure has been a
  gold mine for those of us interested in planet formation because it
  provides vital clues about this process.  This chapter is a review
  of the current state of knowledge about these issues.  It} is
  divided in two parts. The first part (section~\ref{current_dyn}) is
  devoted to the description of the current dynamics in the Kuiper
  belt.  This will be used in section~\ref{problems} to {highlight}
  the properties of the Kuiper belt population that cannot be
  explained by the current dynamical {processes}, but need to be
  understood in the framework of a scenario of primordial evolution of
  the outer Solar System.  In the second part, we will review the
  models that have been proposed so far to explain the various
  puzzling properties of the Kuiper belt.  More precisely,
  section~\ref{edge} will focus on the origin of the outer edge of the
  belt; section~\ref{migration} will describe the effects of the
  migration of Neptune on the orbital structure of the Kuiper belt
  objects {(KBOs)} and section~\ref{depletion} will discuss the origin
  of the mass deficit of trans-Neptunian population. In
  section~\ref{Nice} we will present the consequences on the Kuiper
  belt of a model of the outer Solar System evolution that has been
  recently proposed to explain the orbital architecture of the
  planets{, the Trojan populations of both Jupiter and Neptune,} and
  the origin of the Late Heavy Bombardment of the terrestrial
  planets. A general discussion on the current {state-of-the-art} in
  Kuiper belt modelling will close the paper in
  section~\ref{conclusions}.

\section{Current dynamics in the Kuiper belt}
\label{current_dyn} 

\begin{figure}[t!]
\centerline{\psfig{figure=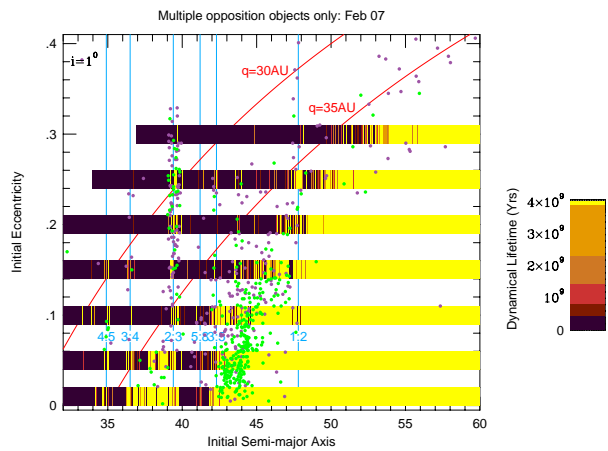,height=6.5cm,angle=0}}
\caption{ The dynamical lifetime for {massless} particles in the
  Kuiper belt derived from 4 billion year integrations ({\it Duncan et
  al.}, 1995; {but extended to $a=60$ AU for this review}).  Each particle
  is represented by a narrow vertical strip of color, the center of
  which is located at the particle's initial eccentricity and
  semi-major axis (the initial orbital inclination for all objects was
  1 degree).  The color of each strip represents the dynamical
  lifetime of the particle, as reported on the scale on the right hand
  side.  For reference, the locations of the important Neptune
  mean-motion resonances are shown in blue and two curves of constant
  perihelion distance, $q$, are shown in red.  The $(a,e)$ elements of
  the Kuiper belt objects with orbits determined over 3 oppositions
  are also shown. Green dots are for $i<4^\circ$, magenta
  dots otherwise. }
\label{Hal} 
\end{figure}

Fig.~\ref{Hal} shows a map of the dynamical lifetime of
trans-Neptunian bodies as a function of their initial semi-major axis
and eccentricity, for an inclination of $1^\circ$ {and with their
orbital ellipses oriented randomly} ({\it Duncan et
  al.}, 1995).  {Additional} maps, referring to different
choices of the initial inclination or different projections on orbital
element space can be found in {{\it Duncan et al.}, 1995 and
  {\it Kuchner et al.}, 2002}.  These maps have been computed
numerically, by simulating the evolution of massless particles from
their initial conditions, under the gravitational perturbations of the
giant planets. The planets were assumed to {be on} their
current orbits {throughout the integrations}. Each particle was
followed until it suffered a close encounter with Neptune.  Objects
encountering Neptune would then evolve in the Scattered disk (see {\it
  chapter} by {\it Gomes et al.}).

In fig.~\ref{Hal}, the colored strips indicate the {timespan} required
for a particle to encounter Neptune, as a function of its initial
semi-major axis and eccentricity.  Strips that are colored yellow
represent objects that survive for the length of the simulation,
$4\times 10^9$ years (the approximate age of the Solar System) without
encountering the planet.  The figure also reports the orbital elements
of the known Kuiper belt objects. Green dots refer to bodies with
inclination $i<4^\circ$, consistent with the low inclination at which
the stability map has been computed. Magenta dots refer to objects with
larger inclination and are plotted only for completeness.

{As the figure shows}, the Kuiper belt has a complex dynamical
structure, although some general trends can be easily
{identified}. {If we denoted the perihelion distance of an
orbit by $q$, and we note that $q=a(1-e)$, where $a$ is the
semi-major axis and $e$ is the eccentricity, Fig.~\ref{Hal} shows
that} most objects with $q\lesssim 35$~AU (in the region $a<40$~AU) or
$q\lesssim 37$--38~AU (in the region with $42<a<50$~AU) are unstable.
This is due to the fact that they pass sufficiently close to Neptune
to be destabilized. It may be surprising that Neptune can destabilize
objects passing at a distance of 5--8~AU, which corresponds to
{$\sim\!10$ times the radius of Neptune's gravitational sphere of
influence or Hill radius.} The instability, in fact, is not due to
close encounters with the planet, but to the overlapping of its outer
mean motion resonances. It is well known that mean motion resonances
become wider at larger eccentricity (see for instance {{\it
Dermott and Murray}, 1983 and also} {\it Morbidelli}, 2002) and that
resonance overlapping produces large scale chaos ({\it Chirikov},
1960). The overlapping of resonances produces a chaotic band whose
extent in perihelion distance away from the planet is proportional to
the planet mass at the 2/7 power ({\it Wisdom}, 1980). To date, the
most extended analytic calculation of the width of the mean motion
resonances with Neptune up to order 50 has been done by D.~Nesvorny,
and the result -in good agreement with the stability boundary observed
in Fig.~\ref{Hal}- is published electronically at
http://www.boulder.swri.edu/{$\sim$}davidn/kbmmr/kbmmr.html.

The semi-major axis of the objects that are above the resonance
overlapping limit evolves by `jumps', passing from the vicinity of 
one resonance  to another, mostly during a conjunction with the
planet. Given that the eccentricity of Neptune's orbit is small, 
the {\it Tisserand parameter} 
$$
T=\frac{a_N}{a} + 2 \sqrt{\frac{a}{a_N}(1-e^2)}\cos i
$$
($a_N$ denoting the semi-major axis of Neptune, $a,e,i$ the
semi-major axis, eccentricity and inclination of the object) is
approximately conserved.  Thus, the eccentricity of the object's orbit
has `jumps' correlated to those of the semi-major axis, and the
perihelion distance remains roughly constant. Consequently, the object
wanders over the $(a,e)$ plane and is effectively a member of the
Scattered {disk.} In conclusion, the boundary between the black
and the colored region in Fig.~\ref{Hal} marks the boundary of the
Scattered disk and has a complicated, fractal structure, which
justifies the use of numerical simulations in order to classify the
objects (see {\it Chapters} by {\it Gladman et al.} and {\it Gomes et al.}).

Not all bodies with $q<35$~AU are unstable, though. The exception is
those objects deep inside low-order mean-motion resonances with
Neptune.  These objects, despite approaching (or even intersecting)
the orbit of Neptune at perihelion, never {get close to the
  planet.  {The reason for this can be understood with a little
algebra.  If a body is in a $k_N$:$k$ resonance with Neptune, the
ratio of its orbital period $P$ to Neptune's $P_N$ is{, by
definition,} equal to $k/k_N$. In this case, and assuming that
the planet is on a quasi-circular orbit and the motions of the particle
and of the planet are co-planar, the angle \begin{equation}
\sigma=k_N\lambda_N-k\lambda+(k-k_N)\varpi, \label{sigma-def}
\end{equation} (where $\lambda_N$ and $\lambda$ denote the mean
longitudes of Neptune and of the object and $\varpi$ is the object's
longitude of perihelion) has a time derivative which is zero on
average, so that it librates around an equilibrium value, say
$\sigma_{\rm stab}$ {(see Fig.~\ref{res-portrait})}.  The radial
distance from the planet's orbit is minimized when the object passes
close to perihelion. Perihelion passage happens when
$\lambda=\varpi$. When this occurs, from (\ref{sigma-def}) we see that
the angular separation between the planet and the object,
${\lambda}_N-\lambda$, is equal to $\sigma/k_N$.  {For small amplitude
librations,} $\sigma\sim\sigma_{\rm stab}$; because $\sigma_{\rm
stab}$ is typically far from 0 {see (Fig.~\ref{res-portrait})}, we
conclude that close encounters cannot occur ({\it Malhotra},
1996). {Conversely, if the body is not in resonance, $\sigma$
circulates (Fig.~\ref{res-portrait}). So, eventually it has to pass
through 0, which brings the object to be in conjuction with Neptune
during its closest approach to the planet's orbit. Thus, close
encounters are possible, if the object's perihelion distance is small
enough.}}

\begin{figure}[t!]
\centerline{\psfig{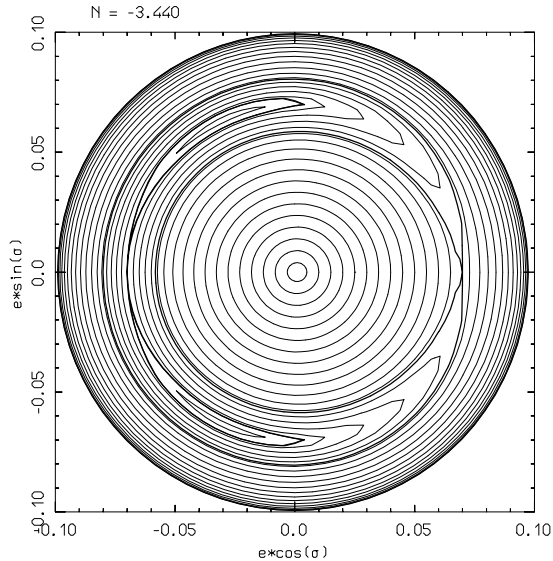}}
\caption{The dynamics in the 1:2 resonance with Neptune, in
  $e\cos\sigma, e\sin\sigma$ coordinates. The motions follow the
  closed curves plotted in the figure. Librations occur on those
  curves that do not enclose the origin of the figure. Notice two
  unstable equilibrium points on the $e\sin\sigma=0$ line. Each
  unstable equilibrium is the origin of a critical curve called {\it
  separatrix}, plotted in bold in the figure. The one with origin at
  the unstable point at $\sigma=0$ separates resonant from not
  resonant orbits, for which $\sigma$ respectively librates or
  circulates. The separatrix {with origin at the unstable point in}
  $\sigma=180^\circ$ delimits two islands of libration around
  each of the asymmetric stable equilibria.}
\label{res-portrait} 
\end{figure}

For most mean motion resonances, $\sigma_{\rm stab}=180^\circ$.
However, {this is not true for the} resonances of type 1:$k$.  In
these resonances, if the eccentricity of the body is not very small,
there are two stable equilibria at $\sigma_{\rm
stab}=180^\circ\pm\delta$, with $\delta\sim 60^\circ$, while
$\sigma=180^\circ$ is an unstable equilibrium ({\it Message}, 1958;
{\it Beaug\'e}, 1994; see Fig.~\ref{res-portrait}).  Thus, bodies with
small amplitudes of libration, necessarily librate asymmetrically in
$\sigma$ relative to the $(0,2{\pi})$ interval. Symmetric librations
are possible only for large amplitude librators.

A detailed exploration of the stability region inside the two main
mean-motion resonances of the Kuiper belt, the 2:3 and 1:2 resonances
with Neptune, has been done in {\it Nesvorny and Roig}, 2000, 2001).
In general, they found that orbits with large amplitude of libration
and moderate to large eccentricities are chaotic, and eventually
escape from the resonance, joining the scattered disk population.
Conversely, orbits with small eccentricity or small libration
amplitude are stable over the age of the Solar System. At large
eccentricity, only asymmetric librations are stable in the 1:2
resonance.

Mean motion resonances are not the only important agent structuring
the dynamics in the Kuiper belt.  In fig.~\ref{Hal}, one can see that
{there is a dark region that extends down to $e=0$ when
  $40<a<42$~AU}.  The instability in this case is due to the presence
of {the secular resonance that occurs when $\dot\varpi \sim
  \dot\varpi_N$}, where $\varpi_N$ is the perihelion longitude of
Neptune. More {generally}, secular resonances occur when the
precession rate of the perihelion or of the longitude of the node of
an object is equal to the mean precession rate of the perihelion or
the node of one of the planets. The secular resonances involving the
perihelion precession rates excite the eccentricities, while those
involving the node precession rates excite the inclinations ({\it
  Williams and Faulkner}, 1981; {\it Morbidelli and Henrard}, 1991).

The location of secular resonances in the Kuiper belt has been
computed in {\it Knezevic et al.} (1991). This work showed that both
the secular resonance with the perihelion and that with the node of
Neptune are present in the $40<a<42$~AU region, for $i<15^\circ$.
Consequently, a low-inclination object in this region undergoes large
variations in orbital eccentricity so that --~even if the initial
eccentricity is zero~-- the perihelion distance eventually decreases
below 35~AU, and the object enters the Scattered disk ({\it Holman and
  Wisdom}, 1993; {\it Morbidelli et al.}, 1995). Conversely, a large
inclination object in the same semi-major axis region is stable.
Indeed, fig.~\ref{Hal} shows that many objects with $i>4^\circ$ (small
dots) are present between 40 and 42~AU. Only large dots, representing
low-inclination objects, are absent.

\begin{figure}[t!]
\centerline{\psfig{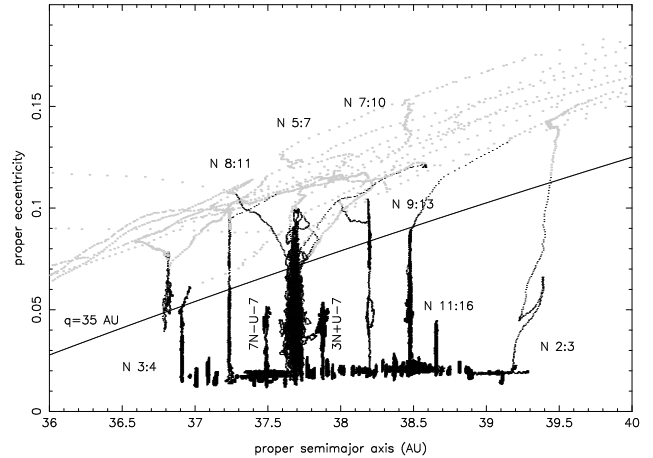}}
\caption{The evolution of objects initially at $e=0.015$ and semi-major axes
  distributed in the 36.5--39.5 AU range. The dots represent the
  proper semi-major axis and the eccentricity of the objects
  --computed by averaging their $a$ and $e$ over 10~My time
  intervals-- over the age of the Solar System.  They are plotted in
  gray after the perihelion has decreased below 32 AU for the first
  time. Labels {$Nk_N$:$k$} denote the {$k_N$:$k$}
  two-body resonances with Neptune.  Labels $k_N$N$+k_U$U$+k$ denote
  the three-body resonances with Uranus and Neptune, corresponding to
  the equality $k_N\dot\lambda_{\rm N}+k_U\dot\lambda_{\rm
    U}+k\dot\lambda=0$.  Reprinted from {\it Nesvorn\'y and
      Roig}, (2001).}
\label{diffusion} 
\end{figure}

{Another} important characteristic revealed by Fig.~\ref{Hal}
is the presence of narrow regions, represented by brown colored bands,
where orbits become Neptune-crossing only after billions of years of
evolution. The nature of these weakly unstable orbits remained
mysterious for several years. Eventually, it was found ({\it Nesvorny
  and Roig}, 2001) that they are, in general, associated either with
high order mean-motion resonances with Neptune (i.e.  resonances for
which the equivalence {$k\dot\lambda =k_N \dot\lambda_N$} holds
only for large values of the integer coefficients $k, k_N$) or
three--body resonances with Uranus and Neptune (which occur when
{$ k\dot\lambda + k_N\dot\lambda_N + k_U \dot\lambda_U=0$}
occurs for some integers $k, k_N$ and $k_U$).

The dynamics of objects in these resonances is chaotic due to the
non--zero eccentricity of the planetary orbits.  The semi-major axis
of the objects remain locked at the corresponding resonant value,
while the eccentricity of their orbits {slowly evolves}. In an
$(a,e)$-diagram like fig.~\ref{diffusion}, each object's evolution
leaves a vertical trace.  This phenomenon is called {\it chaotic
diffusion}.  Eventually the growth of the eccentricity can bring the
diffusing object to decrease the perihelion distance below
$35$~AU. These resonances are too weak to offer an effective
protection against close encounters with Neptune ($\sigma_{\rm
stab}/k_N$ is a small quantity because $k_N$ is large), unlike the low
order resonances considered above.  Thus, once the perihelion distance
becomes too low, the encounters with Neptune start to change the
semi-major axis of the objects, which leave their original resonance
and evolve --~from that moment on~-- in the Scattered disk.

Notice from fig.~\ref{diffusion} that some resonances are so weak
that, despite forcing the resonant objects to diffuse chaotically,
they cannot reach the $q=35$~AU curve within the age of the Solar
System. Therefore, {although these objects are not stable from
  the dynamics point of view, they can be consider that way from the
  astronomical perspective}.

Notice also that chaotic diffusion is effective only for selected
resonances.  The vast majority of the simulated objects are not
affected by any macroscopic diffusion. They preserve their initial
small eccentricity for the entire age of the Solar System. Thus, the
current moderate/large eccentricities and inclinations of most of the
Kuiper belt objects cannot be obtained from primordial circular and
coplanar orbits by dynamical evolution in the framework of the current
orbital configuration of the planetary system.  Likewise, the region
beyond the 1:2 mean-motion resonance with Neptune is totally stable {up
to an eccentricity of $\sim 0.3$}
(Fig.~\ref{Hal}).  {As a result}, the absence of bodies beyond
48 AU cannot be explained by current dynamical instabilities.
Therefore, these (and other) intriguing properties of the Kuiper
belt's structure must, instead, be explained within the framework of
the formation and primordial evolution of the Solar System. These
topics will be treated in the next sections.

\section{Kuiper belt properties acquired during a primordial age}
\label{problems}

From the current dynamical structure of the Kuiper belt, we conclude
that the properties that require an explanation in the framework of
the primordial Solar System evolution are:

\medskip

\noindent{\it i}) The existence of conspicuous populations
of objects in the main mean motion resonances with Neptune (2:3, 3:5,
4:7, 1:2, 2:5, etc.). {The} dynamical analysis presented above
shows that these resonances are stable, but does not explain how and
why objects populated these resonances on orbits with eccentricities
as large as allowed by stability considerations.

\medskip
\noindent{\it ii}) The excitation of the eccentricities in
the classical belt, which we define here as the collection of
non-resonant objects with $42<a<48$~AU {and $q>37$~AU}.  The
median eccentricity of the classical belt is $\sim {0.07}$.  It should
be noted, however, that the upper eccentricity boundary of this
population is set by the long-term orbital stability of the Kuiper
belt (see Fig.~\ref{Hal}), and thus this semi-major axis region could
have contained at some time in the past objects with much larger
eccentricities.  In any case, even if the current median eccentricity
is small, it is nevertheless {much larger (an order of
  magnitude or more) than the one that must have existed} when the
{KBOs} formed. The current dynamics are
stable{, so that}, without additional stirring mechanisms, the
primordial small eccentricities should have been preserved {to
  modern times}.

\medskip
\noindent{\it {iii}}) The peculiar ($a,e$) distribution of the
objects {in the classical belt} (see Fig.~\ref{Hal}).
{In particular, the} population of objects on nearly-circular
orbits {($e\!\lesssim0.05$)} effectively {ends} at about
44~AU, and beyond this {location} the eccentricity tends to
increase with semi-major axis. {If this were simply the
  consequence of an observational bias that favors the discovery of
  objects on orbits with smaller perihelion distances, we would expect that 
  the lower bound of the {$a$--$e$} distribution in the 44--48 AU
{would follow} a curve of constant $q$. {This is not the
case.  Indeed, the eccentricity of this boundary grows more steeply
with semi-major axis than this explaination would predict.}  Thus, the
apparent relative under-density of objects at low eccentricity in the
region $44<a<48$~AU is likely to be a real feature of the Kuiper belt
distribution. This under-density cannot be explained by a lack of
stability {in} this region.}

\medskip
\noindent{\it {iv}}) The outer edge of the classical belt
(Fig.~\ref{Hal}).  This edge appears to be precisely at the location
of the 1:2 mean-motion resonance (MMR) with Neptune. Only large
eccentricity objects, typical of the scattered disk or of the detached
population (see {\it Chapter} by {\it Gladman et al.} for a definition
of these populations) seem to exist beyond this boundary
{(Fig.~\ref{Hal})}. Again, the under density (or absence) of low
eccentricity objects beyond the 1:2 MMR cannot be explained by
observational biases ({\it Trujillo and Brown}, 2001; {\it Allen et
al.}, 2001, 2002; see also {\it Chapter} by {\it Kavelaars et al.}){.}
As Fig.~\ref{Hal} shows, the region beyond the 1:2 resonance looks
stable, even at moderate eccentricity.  So, a primordial distant
population should have remained there.

\begin{figure}[t!]
\centerline{\psfig{figure=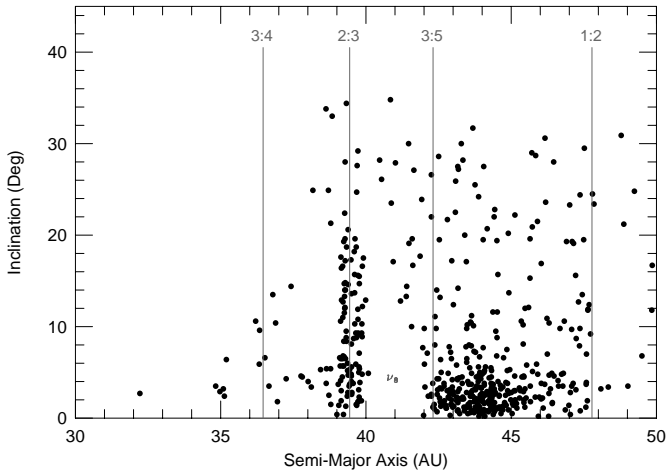,height=7.cm}}
\caption{{The semi-major axis --- inclination distribution of
    all well observed Kuiper belt objects.  The important mean motion
    resonances are also shown.}}
\label{fig_ai} 
\end{figure}

\medskip
\noindent{\it {v}}) The inclination distribution in the
classical belt. The observations {(see Fig.\ref{fig_ai})} show
a {clump} of objects with $i\lesssim 4^\circ${. However,
  there are} also several objects with much larger inclinations, up to
$i\sim 30^\circ$, despite the {fact that an object's
  inclination does not change much in the current Solar System.}
Observational biases definitely enhance the low inclination
{clump} relative to the large inclination population (the
probability of discovery of an object in an ecliptic survey is roughly
proportional to $1/\sin(i)$). However, the {clump} persists
even when the biases are taken into account.  {\it Brown} (2001)
{argued} that the debiased inclination distribution is bimodal and can
be fitted with two Gaussian functions, one with a standard deviation
$\sigma \sim 2^\circ$ for the low-inclination core, and the other with
$\sigma \sim 12^\circ$ for the high inclination population (see also {\it
  Chapter} by {\it Kavelaars et al.}). Since the work of Brown, the
classical population with $i<4^\circ$ is called the `cold population', and
the remaining one is called the `hot population' (see {\it Chapter} by
{\it Gladman et al.} for nomenclature issues).

\medskip
\noindent{\it {vi}}) The correlations between physical properties and
orbital distribution. The cluster of low inclination objects visible
in the ($a,i$) distribution disappears if one selects only objects
with absolute magnitude $H<6$ ({\it Levison and Stern},
2001)\footnote{{The absolute magnitude is the brightness that the
object would have if it were viewed at 1 AU, with the observer at the
Sun. It relates to size by the formula $Log D^2=6.244-0.4 H-Log(p)$,
where $D$ is the diameter in kilometers and $p$ is the albedo.}}. This
implies that intrinsically bright objects are under represented in the
cold population.  {\it Grundy et al.} (2005) have shown that the
objects of the cold population have a larger albedo, on average, than
those of the hot population. Thus, the correlation found by Levison
and Stern implies that the hot population contains bigger objects.
{\it Bernstein et al$.$}~(2004) showed that the hot population has a
shallower $H$ distribution than the cold population, which is
consistent with the absence of the largest objects in the cold belt.
In addition, there is a well known correlation between color and
inclination (see {\it Chapter} by {\it Doressoundiram et al.}).  The
hot population objects show a wide range of colors, from red to gray.
Conversely, the cold population objects are mostly red.  In other
words, the cold population shows a significant deficit of {gray
bodies} relative to the hot population. The differences in physical
properties argue that the cold and the hot populations have different
origins.

\medskip

\noindent{\it {vii}}) The mass deficit of the Kuiper belt. The
current mass of the Kuiper belt is very small{.  Estimates
  range} from 0.01~Earth masses ($M_\oplus$) ({\it Bernstein et al.}, 2004) to
0.1~$M_\oplus$ ({\it Gladman et al.}, 2001). The uncertainty is due
mainly to the conversion from absolute magnitudes to sizes,
{assumptions about} bulk density, and {ambiguities in the size
  distribution (see {\it Chapter} by {\it Petit et al.}).}  Whatever
the exact real total mass, there appears to be a significant mass
deficit (of 2--3 orders of magnitude) with respect to what models say
is needed in order for the KBOs to accrete {\it in situ}. In
particular, {in order to grow the objects that we see within a
  reasonable time ($10^7$--$10^8$~My), the Kuiper belt must have
  consisted} of about 10 to 30~$M_\oplus$ of solid material in a
dynamically cold disk ({\it Stern}, 1996; {\it Stern and Colwell},
1997a, 1997b; {\it Kenyon and Luu}, 1998, 1999a, 1999b; {\it Kenyon
  and Bromley}, 2004a). If most of the Kuiper belt is currently
stable, and therefore objects do not escape, what {depleted
  $\gtrsim\!99.9\%$} of the Kuiper belt primordial mass?

\medskip

All these issues provide us with a large number of clues to understand
what happened in the outer Solar System during the primordial
{era}.  Potentially, the Kuiper belt might teach us more about
the formation of the giant planets than the planets themselves. This
is what makes the Kuiper belt so important and fascinating for
planetary science.

\section{Origin of the outer edge of the Kuiper belt}
\label{edge}

The existence of an outer edge of the Kuiper belt is very
intriguing. Several mechanisms for its origin have been proposed, none
of which have resulted yet in a general consensus {among} the
experts in the field.  These mechanisms can be grouped in three
classes.

\paragraph{Class I: Destroying the distant planetesimal disk}
It has been {argued}  in {\it Brunini and Melita} (2002)
that a Martian mass body residing for 1~Gy on an orbit with $a\sim
60$~AU and $e\sim 0.15$--0.2 could have scattered most of the Kuiper
belt bodies originally in the 50--70~AU range into Neptune-crossing
orbits, leaving this region strongly depleted and dynamically excited.
It might be possible (see {\it Chapter} by {\it Kavelaars et al.})
that the apparent edge at 50~AU is simply the inner edge of such a gap
in the distribution of Kuiper belt bodies.  A main problem with this
scenario is that there are no evident dynamical mechanisms that would
ensure the later removal of the massive body from the system. In other
words, the massive body should still be present, somewhere in the
$\sim 50-70$~AU region.  A Mars-size body with 4\% albedo at 70~AU
would have apparent magnitude brighter than 20.  In addition, its
inclination should be small, both in the scenario where it was
originally a Scattered disk object whose eccentricity (and
inclination) were damped by dynamical friction (as envisioned by
Brunini and Melita), and in {the one} where the body reached
its required heliocentric distance by migrating through the
primordially massive Kuiper belt (see {\it Gomes et al.}, 2004).
Thus, in view of its brightness and small inclination, it is unlikely
that the putative Mars-size body could have escaped detection in the
numerous wide field ecliptic surveys that have been performed up to
now, and in particular in that described in {\it Trujillo and Brown}
{(2003)}.

A second possibility {for destroying the Kuiper belt beyond the
  observed edge} is that the planetesimal disk was truncated by a
  close stellar encounter.  The eccentricities and inclinations of the
  planetesimals resulting from a stellar encounter depend critically
  on $a/D$, where $a$ is the semi-major axis of the planetesimal and
  $D$ is the {closest} heliocentric distance of the stellar encounter
  ({\it Ida et al.}, 2000; {\it Kobayashi and Ida}, 2001).  {An
  encounter with a solar-mass star}   at
  $\sim\!200$~AU would make most of the bodies beyond 50~AU so
  eccentric that they intersect the orbit of Neptune, which would
  eventually produce the observed edge ({\it Melita et al.}, 2002).
  An interesting constraint on the time at which such an encounter
  occurred is set by the existence of the Oort cloud. It was shown in
  {\it Levison et al.}  (2004) that the encounter had to occur much
  earlier than $\sim 10$~My after the formation of Uranus and Neptune,
  otherwise most of the existing Oort cloud would have been ejected to
  interstellar space.  Moreover, many of the planetesimals in the
  Scattered disk at that time would have had their perihelion distance
  lifted beyond Neptune, decoupling them from the planet.  As a
  consequence, the detached population, with {$50\lesssim a\lesssim
  100$}~AU and $40<q<50$~AU, would have had a mass comparable or
  larger than that of the resulting Oort cloud, hardly compatible with
  the few detections of detached objects achieved up to
  now. {Finally, this mechanism predicts an correlation between
  inclination and semi-major axis that is not seen.}

  {A way around the above problems could be achieved if the
  encounter occurred during the first million years of Solar System
  history (Levison et al., 2004).  At this time, the Sun was still in
  its birth cluster making such an encounter likely.
  However, the Kuiper belt objects were presumably not yet fully
  formed ({\it Stern}, 1996; {\it Kenyon and Luu}, 1998), and thus an
  edge to the belt would form at the location of the disk where
  eccentricities are $\sim 0.05$.  Interior to this location}
collisional damping is efficient and accretion can recover
 {from the encounter, beyond this location} the objects rapidly
grind down to dust ({\it Kenyon and Broomley}, 2002).  {If this
  scenario is true, the stellar passage cannot be responsible exciting
  the Kuiper belt because the objects that we observe there did not
  form until much later.}
  
According to the analysis done in {\it Levison et al.} (2004), 
an edge-forming stellar encounter should not be the responsible for the
origin of the peculiar orbit of Sedna {($a\!=\!484$~AU and
  $q\!=\!76$~AU)}, unlike the proposed in {\it Kenyon and
  Broomley}, (2004b).  In fact, such a close encounter would also
produce a relative overabundance of bodies with perihelion distance
similar to that of Sedna but with semi-major axes in the 50--200~AU
range ({\it Morbidelli and Levison}, 2004).  These bodies have never
been discovered, despite {the fact that} they should be easier
to find than Sedna because of their shorter orbital period (see {\it
  Chapter} by {\it Gomes et al.}).

\paragraph{Class II: Forming a bound planetesimal 
  disk from an extended gas-dust disk} In {\it Weidenschilling}
(2003), it was {shown} that the outer edge of the Kuiper belt
{might be} the result of two facts: {\it i)} accretion takes
longer with increasing heliocentric distance and {\it ii)} small
planetesimals drift inwards due to gas drag. {According to
  Weidenschilling's models, this} leads to a steepen of the radial
surface density gradient of solids.  The edge effect is augmented
because, at whatever distance large bodies can form, they capture the
$\sim${meter}-sized bodies spiraling inwards from farther out.
The net result of the process, as shown by numerical modeling in {\it
  Weidenschilling} (2003; see Fig.~\ref{stu}), is the production of an
effective edge, where both the surface density of solid matter and the
mean size of planetesimals decrease sharply with increasing distance.

\begin{figure}[t!]
\centerline{\psfig{figure=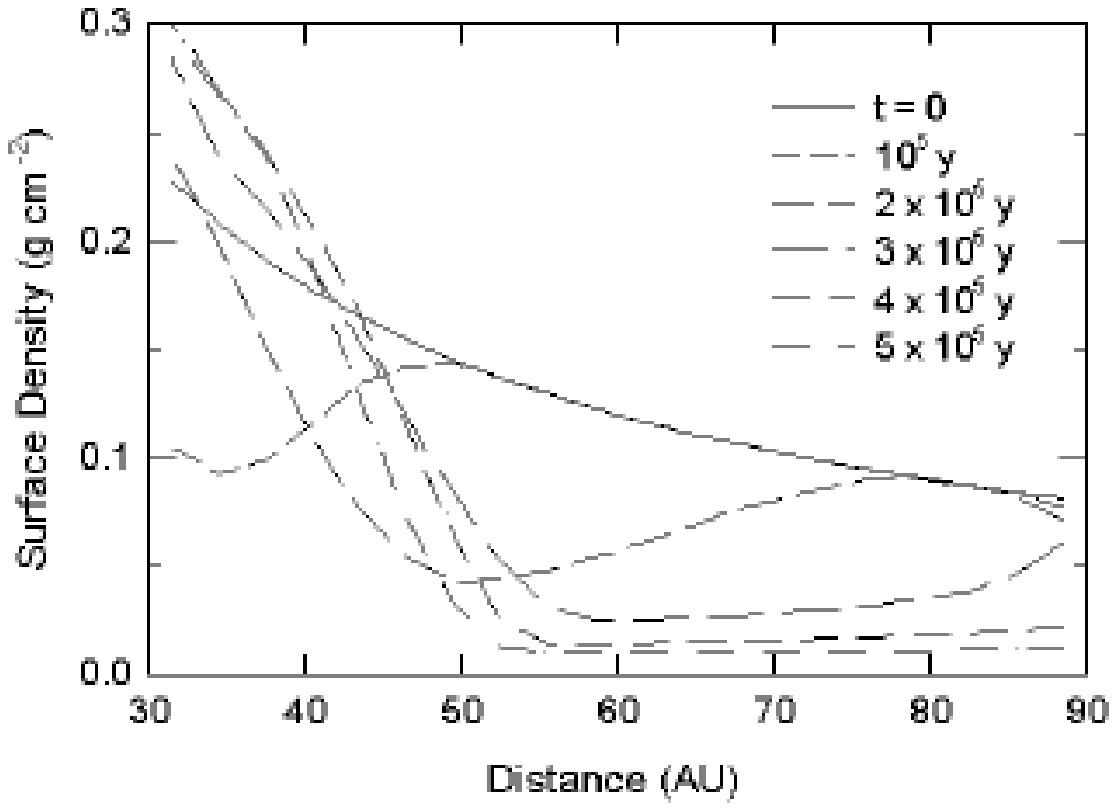,height=6.cm}}
\centerline{\psfig{figure=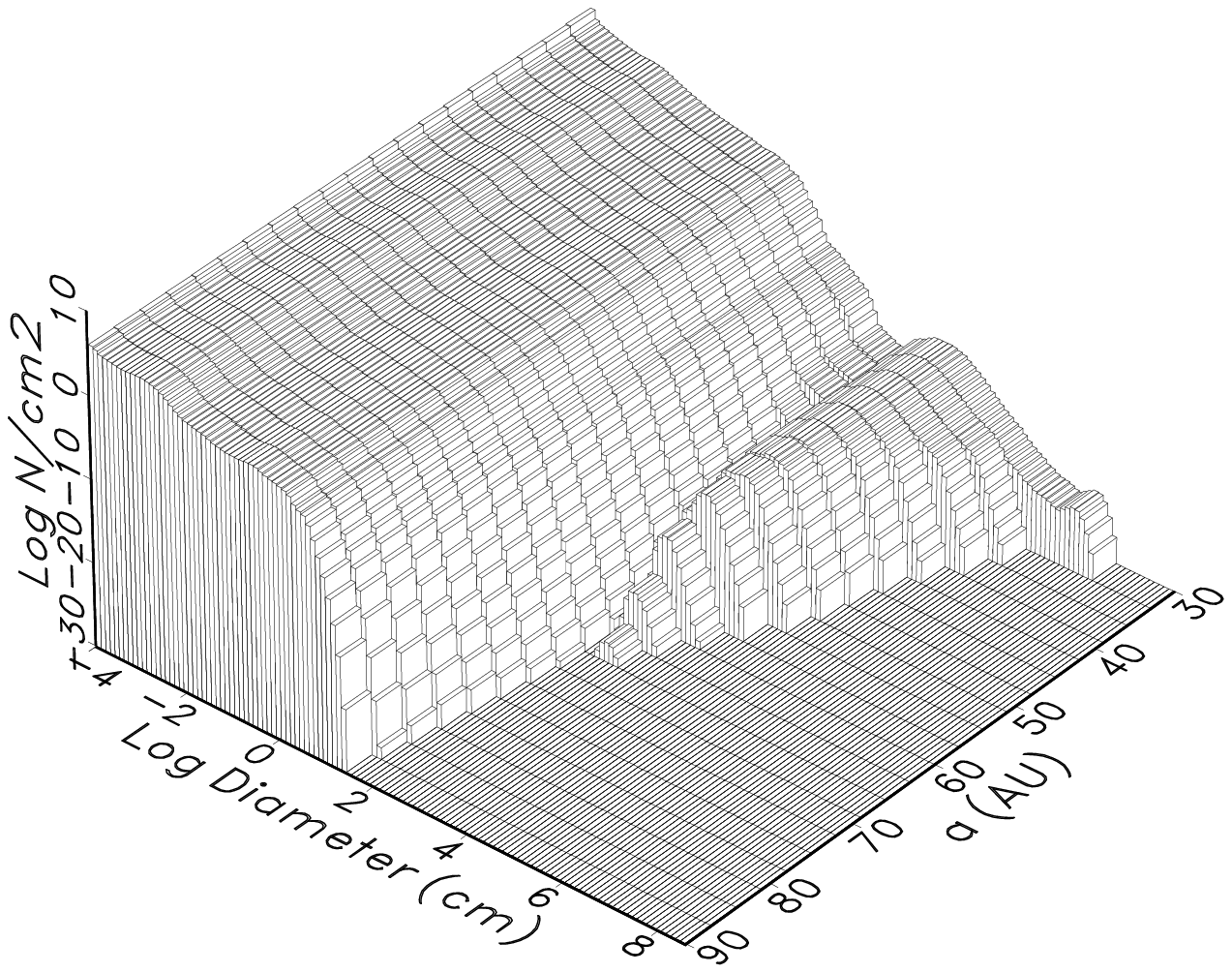,height=5.cm}}
\caption{Top: the time evolution of the surface density of
  solids. Bottom: the size distribution as a function of heliocentric
  distance. From {\it Weidenschilling} (2003).}
\label{stu} 
\end{figure}

A {somewhat similar} scenario has been proposed in {\it Youdin and Shu}
(2002). In their model, planetesimals {formed} by gravitational
instability{, but only in regions of the Solar nebula} where
the local solid/gas ratio was {$\sim\!3$ times that of the Sun
  ({\it Sekiya,} 1983).}  According to the authors, this large ratio
{occurs} because of a radial variation of orbital drift speeds
of millimeter-sized particles induced by gas drag.  This {drift
  also acts to steeping the surface density distribution of the disk
  of solids.  This means that at some point in the nebula, the
  solid/gas ratio falls below the critical value to form
  planetesimals,} so that the resulting planetesimal disk would have
had a natural outer edge.

A third possibility is that planetesimals formed only within a limited
heliocentric distance because of the effect of turbulence. If
turbulence in protoplanetary disks is driven by magneto-rotational
instability (MRI), one can expect that it was particularly strong in
the vicinity of the Sun and at large distances (where solar and
stellar radiation could more easily ionize the gas), while it was
weaker in the central, optically thick region of the nebula, known as
the `dead zone' ({\it Stone et al.}, 1998).  The accretion of
planetesimals should have been inhibited by strong turbulence, because
the latter enhanced the relative velocities of the grains.
Consequently, the planetesimals could have formed only in the dead
zone, with well defined outer (and inner) edge(s).

\paragraph{Class III: Truncating the original gas disk}
The detailed observational investigation of star formation regions has
revealed the existence of many {\it proplyds} (anomalously small
protoplanetary disks). It is believed that these disks were
originally much larger, but in their distant regions the gas was
photo-evaporated by highly energetic radiation emitted by the massive
stars of the cluster ({\it Adams et al.}, 2004). Thus, it has been
proposed that the outer edge of the Kuiper belt reflects the size of
the original Solar System proplyd ({\it Hollenbach et al.}, 2004).

{There is also the theoretical possibility that the disk was born
  small and did not spread out substantially under its own viscous
  evolution ({\it Ruden and Pollack}, 1991). In this case, no truncation
  mechanism is needed. Observations, however, don't show many small
  disks, other than in clusters with massive stars.}

\vskip 10pt
In all the scenarios discussed above, the location of the edge can be adjusted
by tuning the relevant parameters of the corresponding model. In all cases,
however, Neptune played no direct role in the edge formation. In this context,
it is particularly important to remark (as seen in fig.~\ref{Hal}) that the
edge of the Kuiper belt appears to coincide precisely with the location of the
1:2 mean-motion resonance with Neptune. This suggests that, whatever
mechanism formed the edge, the planet was able to adjust the final location of
the outer boundary through gravitational interactions. We will return to this
in section~\ref{pushout}. {Notice that a planetesimal disk
  truncated at $\sim 34$~AU has been recently postulated in order to explain
  the dust distribution in the AU~Mic system ({\it Augereau and
    Beust}, 2006).}

\section{The role of Neptune's migration}
\label{migration}

It was shown in {\it Fernandez and Ip} (1984) that, {in the
  absence of a massive gas disk}, while scattering away the primordial
planetesimals from their neighboring regions, the giant planets had to
migrate in semi-major axis as a consequence of angular momentum
conservation.  Given the configuration of the giant planets in our
Solar System, this migration should have had a general trend (see {\it
  Levison et al.}, 2006 for a review).  Uranus and Neptune have
difficulty ejecting planetesimals onto hyperbolic orbits.  Apart from
the few percent of planetesimals that can be permanently stored in the
Oort cloud or in the Scattered disk, the remaining planetesimals (the
large majority) are eventually scattered inwards, towards Saturn and
Jupiter.  Thus, the ice giants, by reaction, have to move outwards.
Jupiter, on the other hand, eventually ejects from the Solar System
almost all of the planetesimals that it encounters: thus it has to
move inwards. The fate of Saturn is more difficult to predict, {\it a
  priori}. However, {numerical} simulations show that this
planet also moves outwards, although only by a few tenths of an AU for
reasonable disk's masses ({\it Hahn and Malhotra}, 1999; {\it Gomes et
  al.}, 2004).

\begin{figure}[t!]
\centerline{\psfig{figure=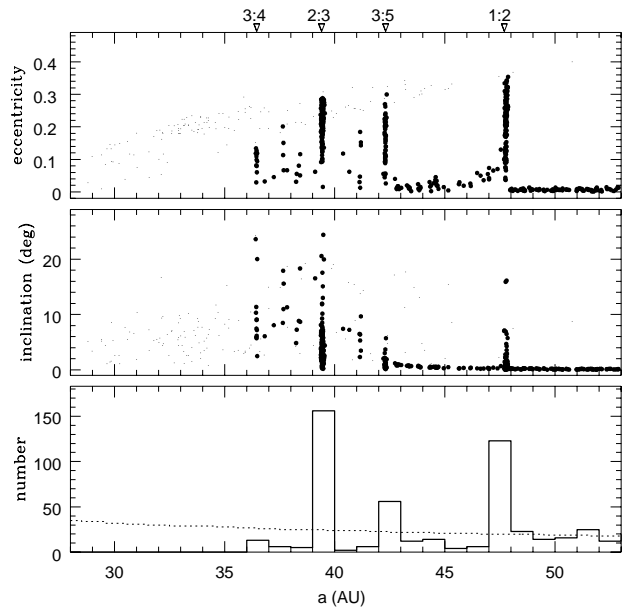,height=8cm}}
\caption{The final distribution of Kuiper belt bodies according to the
  sweeping resonances scenario (courtesy of R.~Malhotra). This simulation was
  done by numerically integrating, over a 200~My time-span, the evolution of
  800 test particles on initially quasi-circular and coplanar orbits. The
  planets are forced to migrate by a quantity $\Delta a$ (equal to $-0.2$~AU
  for Jupiter, 0.8~AU for Saturn, 3~AU for Uranus and 7 AU for Neptune) and
  approach their current orbits exponentially as $a(t)=a_{\infty}-\Delta a
  \exp(-t/4My)$, where $a_\infty$ is the current semi-major axis.  Large solid
  dots represent `surviving' particles (i.e., those that have not suffered any
  planetary close encounters during the integration time); small dots
  represent the `removed' particles at the time of their close encounter with
  a planet (e.g. bodies that entered the Scattered disk and whose evolution
  was not followed further).  In the lowest panel, the solid line is the
  histogram of semi-major axes of the `surviving' particles; the dotted line
  is the initial distribution. The locations of the main mean-motion
  resonances are indicated above the top panel.}
\label{renu} 
\end{figure}

\subsection{The resonance sweeping scenario}
\label{migr-renu}

In {\it Malhotra} (1993{,} 1995) it was realized that, following
Neptune's migration, the mean-motion resonances with Neptune also
migrated outwards, sweeping through the primordial Kuiper belt until
they reached their present positions.  From adiabatic theory (see for
instance {\it Henrard}, 1982), some of the Kuiper belt objects over
which a mean-motion resonance swept, were captured into resonance;
they subsequently followed the resonance through its migration, with
ever increasing eccentricities.  {In fact, it can be shown ({\it
Malhotra}, 1995) that, for a $k_N:k$ resonance, the eccentricity of an
object grows as $$ \Delta e^2= {(k-k_N)\over{k}}\log{{a}\over{a_i}}\ ,
$$ where $a_i$ is the semi-major axis that the object had when it was
captured in resonance and $a$ is its current semi-major axis.}
{This relationship neglects secular effects inside the mean
  motion resonance, that can be important if Neptune's eccentricity is
  not zero and its precession frequencies are comparable to those of
  the resonant particles ({\it Levison and Morbidelli}, 2003)}. This model {can
account} for the existence of the large number of Kuiper belt objects
in the 2:3 mean-motion resonance with Neptune (and also in other
resonances), and {can explain} their large eccentricities (see
fig.~\ref{renu}).  {Assuming} {that all objects
  were captured when their eccentricities were close to zero the above
formula indicates that}, Neptune had to {have migrated} $\sim
7$~AU in order to reproduce quantitatively the observed range of
eccentricities {(up to $\sim 0.3$)} of the resonant bodies.

In {\it Malhotra} (1995), it was also {shown} that the bodies
captured in the 2:3 resonance can acquire large inclinations,
comparable to those of Pluto and other objects.  The mechanisms that
excite the inclination during the capture process have been
investigated in detail in {\it Gomes} (2000), who concluded that,
although large inclinations can be achieved, the resulting proportion
of high inclination {versus} low inclination bodies, as well as
their distribution in the {$e$--$i$} plane, does not reproduce
the observations well. {We will return to this issue in
  section~\ref{sec.gomes}.}

The mechanism of adiabatic capture into resonance requires that
Neptune's migration happened very smoothly. If Neptune had encountered
a significant number of large {bodies,} its jerky migration would have
jeopardized the capture into resonances.  For instance, direct
simulations of Neptune's migration in {\it Hahn and Malhotra} (1999)
--~which modeled the disk with Lunar to Martian-mass planetesimals~--
did not result in any permanent captures.  Adiabatic captures into
resonance can be seen in numerical simulations only if the disk is
modeled using many more planetesimals with smaller masses ({\it
Gomes}, 2003; {\it Gomes et al.}, 2004).
The constraint set by the capture process on the maximum size of the
planetesimals that made up the bulk of the mass in the disk has been
recently estimated in {\it Murray-Clay and Chiang} {(2006).  They
found that resonance capture due to Neptune's migration is efficient
if the bulk of the disk particles was smaller than $\sim\!100\,$km and
the fraction of disk mass in objects with sizes $\gtrsim\!1000\,$km
was less than a few percent.}  {This result appears too severe, because
the results in {\it Gomes}, (2003) and {\it Gomes et al.}, (2004)
show that resonant captures occur in disks entirely
constructed of Pluto-mass objects, although probably with a smaller
efficiency than required in {Murray-Clay and Chiang}'s work.}

{If migration really happened smoothly, {\it Murray-Clay and
    Chiang} (2006) worked out a constraint on the migration
  rate. Remember that there are two islands of libration in the 1:2
  resonance with Neptune (see Fig.~\ref{res-portrait}). If Neptune's
  migration occurs in less than 10~My, they showed that objects
  captured in the 1:2 resonance should preferentially be in the
  trailing island (that wehre $\sigma$ librates around a value
    $\sigma_{\rm stab}>\pi$). Converesely, most of the observed objects are in
  the leading island. This, at first sight, points to a slow Neptune's
  migration. As we will see below, however, resonant objects can be
  captured also from the scattered disk, and Neptune's migration might have
  been very different from what was originally
envisioned. Thus{,} it is
  unclear which kind of constraint is provided by the internal
  distribution of the 1:2 resonant objects.}

As shown in Fig.~\ref{renu}, if the resonance sweeping scenario can
explain the existence of the resonant populations, it cannot explain
the orbital distribution in the classical belt, between 40 and
48~AU{, nor the mass depletion of the Kuiper belt}.  The
eccentricity excitation and, in particular, the inclination excitation
obtained in the simulation in that region are far too small compared
to those inferred from the observations. Thus, {\it Hahn and Malhotra}
(2005) suggested that resonance sweeping occurred after that some
perturbation excited{, and perhaps depleted,} the planetesimal
disk. Thus, the eccentricity and inclination distribution in the
classical belt would not have been sculpted by the sweeping process,
but would be the relic of such previous excitation mechanism(s). A
similar conclusion was reached recently by {\it Lykawka and Mukai}
(2007a), who found that the populations of objects in distant mean
motion resonances with Neptune (i.e. with $a\!>50\!$~AU) and their
eccentricity-inclination-libration amplitude distributions can be
explained by resonance sweeping only if the disk in the 40-48~AU
region was already pre-excited in both $e$ and $i$. 
Possible mechanisms of excitation and their problems will be briefly
discussed in sect.~\ref{depletion}.

\subsection{The origin of the hot population}
\label{sec.gomes}

{The observation that the largest objects in the hot population
  are bigger than those in the cold population lead {\it Levison and
    Stern,} (2001) to suggest that the hot population formed closer to
  the Sun and was transported outward during the final stages of
  planet formation.  {\it Gomes} (2003) showed that the simple
  migration of Neptune described in the previous section could accomplish
  this feat.}

{In particular, {\it Gomes} (2003) studied the migration of the
  giant planets through a disk represented by $10,000$ particles, a
  much larger number than had previously been attempted.}  In Gomes'
simulations, during its migration Neptune scattered the planetesimals
and formed a massive Scattered disk.  Some of the scattered bodies
decoupled from the planet, decreasing their eccentricities through
interactions with some secular or mean-motion resonance (see {\it
  Chapter} by {\it Gomes et al.} for a detailed discussion of how
resonances can decrease the eccentricities). If Neptune were not
migrating, the decoupled phases would have been transient. In fact,
the dynamics are reversible, so that the eccentricity would have
eventually increased back to Neptune-crossing values.  However,
Neptune's migration broke the reversibility, and some of the decoupled
bodies managed to escape from the resonances and remained permanently
trapped in the Kuiper belt.  As shown in fig.~\ref{gomes}, the current
Kuiper belt would therefore be the result of the superposition in
$(a,e)$-space of these bodies with the local population, originally
formed beyond 30~AU.

The local population stayed dynamically cold, in particular in
inclination, because its objects were only moderately excited by
the resonance sweeping mechanism, as in fig.~\ref{renu}.  Conversely,
the population captured from the scattered disk had a much more
extended inclination distribution, for two reasons:{\it(i)} the
inclinations got excited during the scattered disk phase before
capture and {\it (ii)} there was a dynamical bias in favor of the
capture of high-inclination bodies, because at large-$i$ the ability
of mean motion resonances to decrease the eccentricity is enhanced
(see {\it Chapter} by {\it Gomes et al.}).  Thus,  in Gomes model the
current cold and hot populations should be identified respectively
with the local population and with the population trapped
from the scattered disk.

\begin{figure}[t]
\centerline{\psfig{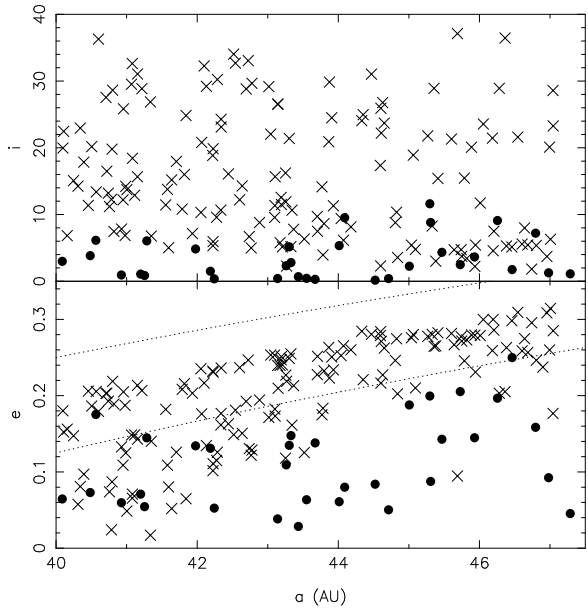}}
\vspace*{-.3cm}
\caption{The orbital distribution in the classical belt according to the 
  simulations in {\it Gomes} (2003). The dots denote the population that formed
  locally, which is only moderately dynamically excited. The crosses denote
  the bodies that were originally inside 30~AU. Therefore, the resulting
  Kuiper belt population is the superposition of a dynamically cold population
  and a dynamically hot population, which gives a bimodal inclination
  distribution.  The dotted curves in the
  eccentricity vs.  semi-major axis plot correspond to $q=30$~AU and
  $q=35$~AU.} \vspace{0.3cm}
\label{gomes}
\end{figure}

This scenario is appealing because, assuming that the bodies' color
varied in the primordial disk with heliocentric distance, it
qualitatively explains why the scattered objects and hot classical
belt objects --~which mostly come from regions inside $\sim$30 AU~--
appear to have similar color distributions, while the cold classical
objects --~the only ones that actually formed in the trans-Neptunian
region~-- have a different distribution. Similarly, assuming that
{at the time of Neptune's migration} the maximum size of the
objects was a decreasing function of {their initial}
heliocentric distance, the scenario also explains why the biggest
Kuiper belt objects are all in the hot population.

As Fig.~\ref{gomes} shows, there may be some quantitative problems in
  the reproduction of the orbital distribution of the hot Kuiper belt
  in Gomes scenario (for instance the perihelion distances appear to
  be somewhat too low). However, an issue of principle concerns 
 the relative weight between the hot and cold
populations. In Gomes's simulations, only a fraction of a percent of
the original scattered disk remained trapped in the hot population. On
the other hand, the cold population was not depleted by the resonance
sweeping, so that it retained most of the original objects. Thus, if
the local population was similar in size distribution and number
density to the planetesimal disk from which the scattered disk was
extracted, it should outnumber the hot population by a huge factor (of
$\sim 1000$).  So, in order to obtain a final inclination distribution
that quantitatively reproduces the debiased inclination distribution
of {\it Brown} (2001), Gomes had to scale down the number of objects
in the cold population by an appropriate factor, assuming that some
mechanism, not included in the simulation, caused a decimation, and
hence a mass depletion, of the local population. These mechanisms are
reviewed in the next section.

Before concluding this section, we {note} that the work by {\it
  Gomes} (2003) also have important implications for the origin of the
detached population. This issue is addressed in detail in the {\it
  Chapter} by {\it Gomes et al.} and therefore we do not discuss it
here.

\section{The mass deficit problem}
\label{depletion}

{As we described in section~\ref{problems}, the Kuiper belt
  only contains roughly 0.1\% of the mass that is required to grow the
  objects that we see.  So, the natural question is what happened to
  all that mass.  We refer to this issue as the `mass deficit
  problem.'  We now review ideas that are currently in the
  literature.}

\subsection{{Mass removal}}
\label{removal}

Two general scenarios have been proposed for the mass depletion: {\it
  (i)} a strong dynamical excitation of the Kuiper belt, which caused
the ejection of most of the bodies from the Kuiper belt to the
Neptune-crossing region, and {\it (ii)} the collisional comminution of
most of the mass of the Kuiper belt into dust.  {We start our
  discussion with (i).}

Because dynamics are size-independent, a dynamical depletion scenario
requires that the primordial population in the Kuiper belt had a size
distribution similar to the one that (currently) exists, but
with a number of objects at each size multiplied by the ratio between
the primordial mass and the current mass. {Remember that, in the
current Solar System configuration, most of the Kuiper belt is stable,
so that dynamical {erosion cannot signficantly reduce} the
total number of objects.} The idea is{, therefore,} that some
perturbation{, that acted in the past and is no-longer at work,}
strongly excited the orbital distribution of the Kuiper belt
population. Most of the original objects acquired Neptune-crossing
eccentricities, so that they were subsequently eliminated by the
scattering action of the planets. Only a small fraction of the
original population, corresponding to the surviving mass fraction,
remained in the Kuiper belt on excited orbits like those of the
observed objects. Thus, this scenario aims at explaining at the same
time the mass depletion of the Kuiper belt and its orbital excitation.

A first dynamical depletion mechanism was proposed in {\it Morbidelli
  and Valsecchi} (1997) and later revisited in {\it Petit et al.}
(2001). This mechanisms invokes the existence of a planetary embryo,
with mass comparable to that of Mars or the Earth, in the scattered
disk for $\sim 10^8$~y.  Another mechanism was proposed by {\it
  Nagasawa et al.} (2000) and invokes the sweeping of secular
resonances through the Kuiper belt during the dispersion of the
primordial gas disk.

The problem with the dynamical depletion scenario, which was not
immediately recognized, is that the ejection of a massive population
of objects from the Kuiper belt to the Neptune-crossing region would
{cause Neptune to migrate into the Kuiper belt.  After all,
  this scenario invokes a $\sim\!15\,M_\oplus$ object to remove
  $\gg15\,M_\oplus$ of disk material and angular momentum must be
  conserved.}  For instance, revisiting the {\it Petit et al$.$}
(2001) work with simulations that account for the effect of the
planetesimals on the dynamics of the massive bodies, {\it Gomes et
  al$.$} {(2004)} showed that even a {disk containing
  $\sim\!4\,M_\oplus$ of material between 40 and 50~AU drives Neptune
  beyond 30~AU.  This is much} less than the mass required
(10--30~$M_\oplus$) by models of the accretion of Kuiper belt bodies
({\it Stern and Colwell}, 1997a; {\it Kenyon and Luu}, 1999b).  

{The} sole possibility for a {viable dynamical model of
  Kuiper belt depletion is if} the objects were kicked directly to
hyperbolic or Jupiter-crossing orbits and were eliminated without
interacting with Neptune.  Only the passage of a star through the
Kuiper belt seems to be capable of such an extreme excitation ({\it
  Kobayashi et al.}, 2005). However, the cold Kuiper belt would not
survive in this case.

We note in passing that, {even if we ignore} the problem of
Neptune's migration, massive embryos or secular resonance sweeping
{are probably} not able to reproduce the inclination
distribution observed in the Kuiper belt. Thus,  these mechanisms
 are unlikely to be an alternative to the scenario
  proposed in {\it Gomes} (2003) for producing the hot classical belt.
 Consequently, the idea of
{\it Hahn and Malhotra} (2005) and {\it Lykawka and Mukai} (2007a)
that the classical belt acquired its current excitation before 
Neptune's migration is not supported, so far, by an
appropriate excitation mechanism.   

The collisional grinding scenario was proposed in {\it Stern and
  Colwell} (1997b) {and} {\it Davis and Farinella} (1997,
1998){,} and then pursued in {\it Kenyon and Luu} (1999a) and
{\it Kenyon and Bromley} (2002, 2004a). It is reviewed in detail in
the {\it Chapter} by {\it Kenyon et al.}.  In essence, a massive
Kuiper belt with large eccentricities and inclinations would
experience {very} intense collisional {grinding}.
Consequently, most of the mass originally in bodies smaller than
several tens of kilometers could be comminuted into dust, and then
evacuated by radiation pressure and Poynting-Robertson drag{.
  This would lead to a substantial depletion in mass.}

To work, the collisional erosion scenario requires that {two}
essential conditions {be} fulfilled. First, it requires a peculiar
primordial size distribution, such that all of the missing mass was
contained in {small, easy-to-break} objects, while the
number of large objects was essentially identical to that in the
current population.  Some models support the existence of such a size
distribution at the end of the accretion phase ({\it Kenyon and Luu},
1998, 1999b).  However, there are several arguments in favor of a
{completely} different size distribution in the planetesimal
disk.  The collisional formation of the Pluto--Charon binary ({\it
Canup}, 2005) and of the 2003~EL$_{61}$ family ({\it Barkume et al.},
2006), the capture of Triton onto a satellite orbit around Neptune
({\it Agnor and Hamilton}, 2006){, and the fact that the Eris,
the largest known Kuiper belt object, is} in the detached population
({\it Brown et al.}, 2005), suggest that the number of big bodies was
much larger in the past, with as many as 1,000 Pluto-sized objects
({\it Stern}, 1991). {Moreover, we have
seen above that the mechanism of {\it Gomes} (2003) for the origin of
the hot population also requires a disk's size distribution with $\sim
1,000$ times more large objects than currently present in the Kuiper
belt.}  Finally, {\it Charnoz and Morbidelli} (2007) showed that, if
the size distribution required for collisional grinding in the Kuiper
belt is assumed for the entire planetesimal disk (5-50 AU), the Oort
cloud and the scattered disk would not contain {enough
comet-size objects to supply the observed} fluxes of long-period and
Jupiter-family comets: the cometesimals would have been destroyed
before being stored in the comet reservoirs {(also see {\it
Stern and Weissman}~2001)}.  So, to fulfill all these constraints and
still have an effective collisional grinding in the Kuiper belt, one
has to assume that the size distribution were totally different in the
{region of the proto-planet disk swept by Neptune and 
in the region of the disk that became the Kuiper belt.}  This,
{\it a priori}, seems unlikely, given the proximity between the two
regions; however, we will come back to this in
sect.~\ref{conclusions}.

The {second} essential condition for substantial collisional grinding
is that the energy of collisions is larger than the energy required
for disruption of the targets. Thus, either the KBOs are extremely
weak (the successful simulations in {\it Kenyon and Bromley} (2004)
had to assume a specific energy for disruption that is at least an
order of magnitude lower than predicted by the
{hydrodynamical{, in particular smooth-particle
hydrodynamical (SPH),} simulations of fragmentation by} {\it Benz and
Asphaug}, 1999), or the massive primordial Kuiper belt had a large
dynamical excitation, with $e\sim 0.25$ and/or $i\sim 7^\circ$ (as
assumed in {\it Stern and Colwell}, 1997b). However, {if, as we
argued above, the hot population was implanted in the Kuiper belt via
the low efficiency process of {\it Gomes}, (2003), then it was never
very massive and would not have had much effect on the collisional
evolution of the cold population.  Thus, the cold population must have
ground itself down.  This is unlikely because the} excitation of the
cold-population is significantly smaller than the required values
reported above. There is the possibility that the collisional erosion
of the cold belt was due to the high-velocity bombardment by
projectiles in the scattered disk. The scattered disk was initially
massive, but its dynamical decay was probably too fast ($\sim 100$~My,
see {\it Duncan and Levison}, 1997). The collisional action of the
scattered disk onto the cold belt was included in {\it Charnoz and
Morbidelli} (2007) but turned out to be a minor contribution.}

\subsection{Pushing out the Kuiper belt}
\label{pushout}

Given the problems explained just above, an alternative way of
solve the mass deficit problem was proposed in {\it Levison and
  Morbidelli} (2003).  In this scenario, the primordial edge of the
massive protoplanetary disk was somewhere around 30--35~AU and the
{\it entire} Kuiper belt population --~not only the hot component as
in {\it Gomes} (2003)~-- formed within this limit and was transported
to its current location during Neptune's migration. The transport
process for the cold population had to be different from the one found
in {\it Gomes} (2003) for the hot population (but still work in
parallel with it), because the inclinations of the hot population were
excited, while those of the cold population were not.

In the framework of the classical migration scenario ({\it Malhotra},
1995; {\it Gomes et al.}, 2004), the mechanism proposed in {\it
  Levison and Morbidelli} (2003) was the following: the cold
population bodies were initially trapped in the 1:2 resonance with
Neptune; then, as they were transported outwards by the resonance,
they were progressively released due to the non-smoothness of the
planetary migration.  In the standard adiabatic migration scenario
({\it Malhotra}, 1995), there would be a resulting correlation between
the eccentricity and the semi-major axis of the released bodies.
However, this correlation was broken by a secular resonance embedded
in the 1:2 mean-motion resonance. This secular resonance was generated
{by the objects in the resonance, themselves. In particular,
  unlike previous studies of migration, Levison and Morbidelli
  included the mass of the objects in the resonance, which modified
  the precession rate of Neptune's orbit}.

Simulations of this process matched the observed $(a,e)$ distribution
of the cold population fairly well, while the initially small
inclinations were only very moderately perturbed.  In this scenario,
the small mass of the current cold population is simply due to the
fact that only a small fraction of the massive disk population was
initially trapped in the 1:2 resonance and then released on stable
non-resonant orbits.  The final position of Neptune would simply
reflect the primitive truncation of the protoplanetary disk
{(see {\it Gomes et al.,} 2004 for a more detailed
  discussion)}.  Most {importantly}, this model explains why
the current edge of the Kuiper belt is at the 1:2 mean-motion
resonance with Neptune, despite the fact that none of the mechanisms
proposed for the truncation of the planetesimal disk involves Neptune
in a direct way (see section~\ref{edge}). The location of the edge was
modified by the migration of Neptune, via the migration of the 1:2
resonance.

On the flip side, the model in {\it Levison and Morbidelli} 2003)
re-opened the problem of the origin of the different physical
properties of the cold and hot populations, because both would have
originated within 35~AU, although in somewhat different parts of the
disk. Moreover, {\it Lykawka and Mukai} (2007a) showed that this model
cannot reproduce the low-to-moderate inclination objects in the
distant (i.e. beyond 50~AU) high-order mean motion resonances with
Neptune.

\section{Effects of a Dynamical Instability in the Orbits of Uranus 
and Neptune}
\label{Nice}

The models reviewed in the previous sections assume that Neptune
migrated outward on a {nearly}-circular orbit. However, a
substantially different model of the evolution of the giant planets
has been recently proposed.  This model --often called the `Nice
model'-- reproduces, for the first time, the orbital architecture of
the giant planet system (orbital separations, eccentricities,
inclinations; {\it Tsiganis et al.}, 2005) and the capture of the
Trojan populations of Jupiter ({\it Morbidelli et al.}, 2005) and
Neptune ({\it Tsiganis et al.}, 2005; {\it Sheppard and Trujillo},
2006).  {It also naturally supplies a trigger for the Late Heavy
  Bombardment (LHB) of the terrestrial planets ({\it Gomes et al.},
  2005), and quantitatively reproduces most of the LHB's
  characteristics.}

\begin{figure}[t!]
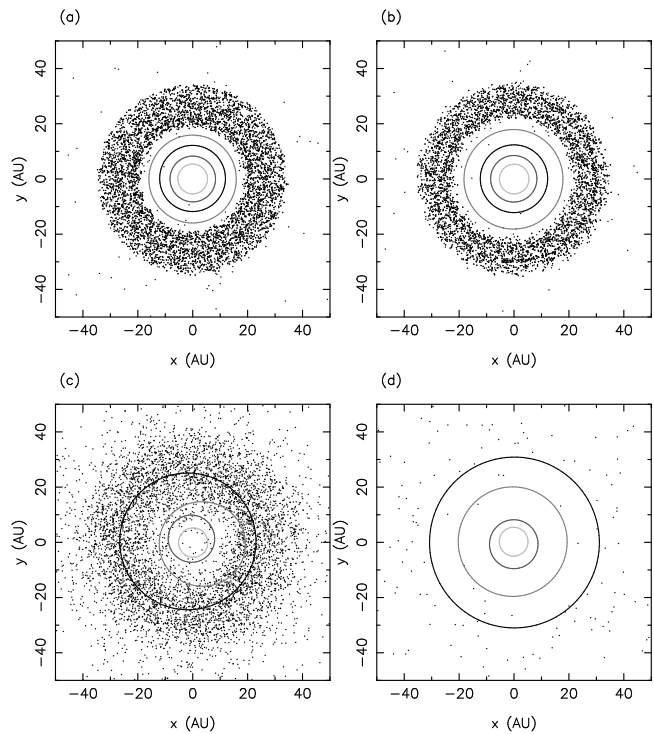

\centerline{\psfig{figure=LHBxy0101.ps,height=4.8cm}\psfig{figure=LHBxy0880.ps,height=4.8cm}}
\centerline{\psfig{figure=LHBxy0883.ps,height=4.8cm}\psfig{figure=LHBxy1200.ps,height=4.8cm}}\vspace*{-.3cm}
\caption{{Solar System evolution in the Nice model. (a): at a time
  close to the beginning of the evolution. The orbits of the giant
  planets (concentric circles) are very close to each other and are
  quasi-circular. They are surrounded by a disk of planetesimals,
  whose inner edge is due to the perturbations from the planets and
  the outer edge is assumed {to be} at 34~AU. (b):
  immediately before the great instability. Saturn is about crossing the
  1:2 resonance with Jupiter. (c): at the {time of the
  instability}. Notice that the orbits of the planets have become
  eccentric and now {penetrate the} planetesimal disk. (d):
  after the LHB. The planets are parked
  on orbits very similar (in terms of separation{,
eccentricity, and inclination}) to
  their current ones. The massive planetesimal disk has been
  destroyed. Only a small fraction of the planetesimals remain in the
  system on orbits typical of the scattered disk{, Kuiper belt,
and other small body reservoirs}.}
  From {\it Gomes et al.} (2005).}
\vspace{0.3cm}
\label{LHB}
\end{figure}

In the Nice model, the giant planets are assumed to be initially on
nearly-circular and coplanar orbits, with orbital separations
significantly smaller than the {ones currently observed}.  More
precisely, the giant planet system is assumed {to lie in the
  region} from $\sim 5.5$~AU to $\sim 14$~AU, and Saturn is assumed to
be closer to Jupiter than their mutual 1:2 MMR. A planetesimal disk is
assumed to exist beyond the orbits of the giant planets, on orbits
whose dynamical lifetime is at least 3~My (the supposed lifetime of
the gas-disk).  The outer edge of the planetesimal disk is assumed
{to lie} at $\sim 34$~AU and the total mass is $\sim\!35
M_\oplus$ {(see Fig.~\ref{LHB}a)}.

With the above configuration, the planetesimals at the inner edge of
the disk {evolve onto Neptune-scattering} orbits on a timescale of a
few {million years}.  Consequently, the migration of the giant planets
proceeds at very slow rate, governed by the slow planetesimal escape
rate from the disk.  Because the planetary system would be stable in
absence of interactions with the planetesimals, this slow migration
continues for a long time, slightly damping out {as the unstable disk
particles are removed from the system} (Fig.~\ref{LHB}). After a long
time, ranging from 350 My to 1.1~Gy in the simulations of {\it Gomes
et al.} (2005) --- which is consistent with the timing of the LHB,
approximately 650~My after planet formation --- Jupiter and Saturn
eventually cross their mutual 1:2 mean-motion resonance
{(Fig.~\ref{LHB}b)}.  This resonance crossing excites their
eccentricities to values slightly larger than those currently
observed. The small jump in Jupiter's and Saturn's eccentricities
drives up the eccentricities of Uranus and Neptune{, however}.  The
ice giant's orbits become chaotic and start to approach each other.
Thus, a short phase of encounters follows the resonance-crossing
event.  Consequently, both ice giants are scattered outward, onto
large eccentricity orbits ($e\sim$~0.3--0.4) that penetrate deeply
into the disk {(Fig.~\ref{LHB}c)}. This destabilizes the full
planetesimal disk and disk particles are scattered all over the Solar
System.  The eccentricities of Uranus and Neptune and --to a lesser
extent-- of Jupiter and Saturn, are damped on a timescale of a few My
due to the dynamical friction exerted by the planetesimals. Thus, the
planets decouple from each other, and the phase of mutual encounters
rapidly ends. During and after the eccentricity damping phase, the
giant planets continue their radial migration, and eventually reach
final orbits when most of the disk has been eliminated
(Fig.~\ref{LHB}d).

The temporary large eccentricity phase of Neptune opens a new degree
of freedom for explaining the orbital structure of the Kuiper belt.
The {new key feature to the dynamics} is that, when Neptune's
orbit is eccentric, the full $(a,e)$ region up to the location of the
1:2 resonance with the planet is chaotic, even {for small
  eccentricities}. This allows us to envision the following scenario.
We assume, in agreement with several of the simulations of the Nice
model, that the large eccentricity phase of Neptune is achieved when
the planet has a semi-major axis of $\sim 28$~AU, after its last
encounter with Uranus. In this case, a large portion of the current
Kuiper belt is already interior to the location of the 1:2 resonance
with Neptune. Thus, it is unstable, and can be invaded by objects
coming from within the outer boundary of the disk (i.e.
${\lesssim}\!34$~AU).  When the eccentricity of Neptune damps
out, the mechanism for the onset of chaos disappears. The Kuiper belt
becomes stable, and the objects that happen to be there at that time
remain trapped for the eternity.  Given that the invasion of the
particles is fast and the damping of Neptune's eccentricity is also
rapid, there is probably not enough time to excite significantly the
particles' orbital inclinations {if} Neptune's inclination is
also small.  Therefore, we expect that this mechanism may be able to
explain the observed cold population. The hot population is then
captured later, when Neptune is migrating up to its final orbit on a
low-eccentricity orbit, as in {\it Gomes} (2003).

\begin{figure}[t!]
\centerline{\psfig{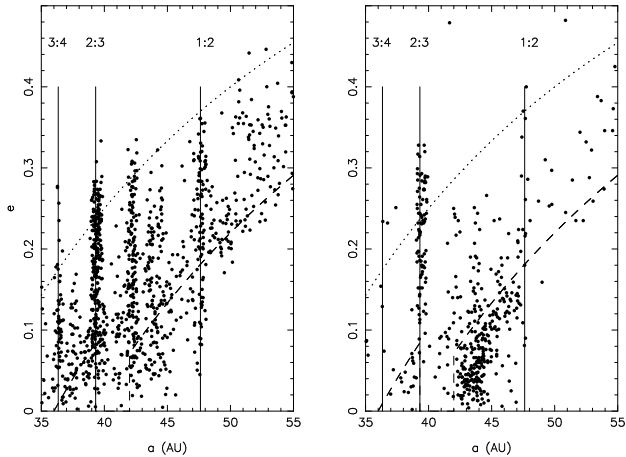}}
\vspace*{-.3cm}
\caption{The distribution of semi major axes
and eccentricities in the Kuiper belt. Left panel: result of a simulation
based on the Nice model. Right panel: the observed
distribution (3 oppositions objects only).  The vertical solid lines mark
the main resonance with Neptune. The dotted curve denotes perihelion distance
equal to 30~AU and the dashed curve delimits the region above which only high
inclination objects or resonant objects can be stable over the age of the
Solar System.}
\vspace{0.3cm}
\label{Nice-ae}
\end{figure}

The numerical simulations of this {process have} been
  presented in {\it Levison et al{$.$}} (2007).  Fig.~\ref{Nice-ae}
compares with the observations the semi-major axis vs. eccentricity
distribution resulting from one of the simulations, 1~Gy after the
giant planet instability. The population of quasi-circular objects at
low inclination extends to $\sim 45$~AU, in nice agreement with the
observations.  The deficit of low eccentricity objects between 45 and
48~AU is reproduced, and the outer edge of the classical belt is at
the final location of the 1:2 MMR with Neptune as observed.  Moreover,
{the real Kuiper belt shows a population
  of objects {with $q\!\sim\!40\,$AU beyond the 1:2 resonance with
  Neptune,} which are
  known to be stable ({\it Emel\'yanenko et al$.$} 2003).  This
  population has been known by several names in the literature, which
  include the fossilized scattered disk, the extended scattered disk,
  and the detached population (see {\it Chapter} by {\it Gladman et
    al.}).  This model reproduces this population quite well.}

{Three} main differences are also noticeable, though. {\it (i)}
All the mean motion resonances are overpopulated relative to the
classical belt. This is probably the consequence of the fact that in
the simulations the migration of Neptune's orbit and its eccentricity
damping were forced smoothly, through fake analytic terms of the
equations of motion. As we said above, a migration with some
stochastic component (due to the encounters with massive objects in
the disk) would have produced fewer surviving bodies in the
resonances. {\it (ii)} The region above the long-dashed curve is
overpopulated in the simulation. The curve represents approximately
the boundary between the stable (yellow) and the unstable (black)
regions in Fig.~\ref{Hal}. Thus, if the final orbits of the giant
planets were exactly the same as the real ones and the simulations
were extended for the age of the Solar System, most of the population
above the curve would be depleted as a consequence of chaotic
dynamics.  {{\it (iii)} The cold Kuiper belt has eccentricities
  that are slightly too large. The median eccentricity of the real
  objects with $42\!<\!a\!<\!48$~AU and $q\!>\!37$~AU is 0.07, while
  the model produces a value of $0.10$.}

\begin{figure}[t!]
\centerline{\psfig{figure=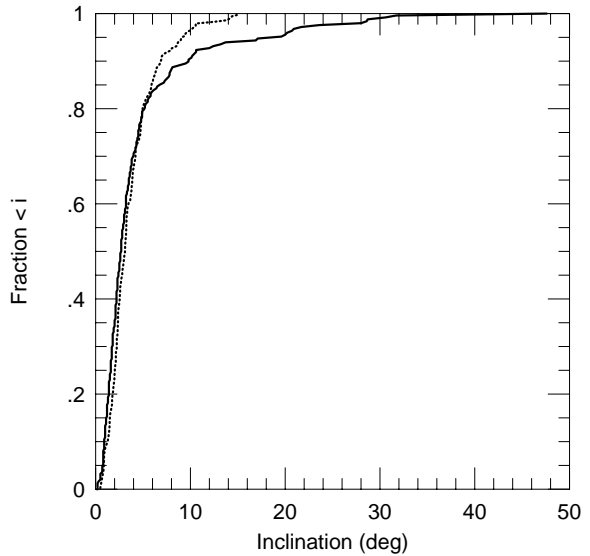,height=7.5cm}}
\vspace*{-.3cm}
\caption{The cumulative inclination distribution of the observed 
  classical belt objects (solid curve) and that expected from the
  result of our simulation, once the observational biases are taken
  into account (dotted curve).}
\vspace{0.3cm}
\label{Nice-i}
\end{figure}

Figure~\ref{Nice-i} shows the cumulative inclination distribution of
objects trapped in the classical belt at the end of a simulation and
compares it with the observed distribution. For the comparison to be
meaningful, the simulated distribution was run through a survey
bias-calculator, following the approach of {\it Brown} (2001). The
{two curves {are very similar.  Indeed, they are almost
indistinguishable for inclinations less then}} 6 degrees.  This means
that the cold population, and the inclination distribution within it,
has been correctly reproduced, as well as the distribution in the
lower part of the hot population (i.e.  that with
$4^\circ<i<10^\circ$). We remark, however, a dearth of large
inclination objects. This deficit is intriguing and unexplained, in
particular given that the raw simulations of the Nice model (namely
those in which the planets are not forced to migrate, but are let free
to respond to their interactions with {massive} planetesimals) produce
objects captured in the classical belt or in the detached population
with inclinations up to $50^\circ$ (see Fig.~3 in the {\it
  Chapter} by {\it Gomes et al.}). In general, we would expect an
inclination distribution in the hot population that is equivalent to
that of {\it Gomes} (2003), or even more excited. In fact, as pointed
out in {\it Lykawka and Mukai} (2007b), the inclinations in the
scattered disk, from which the hot population is derived, are
{restricted to be less than} $\sim 40^\circ$ by the
conservation of the Tisserand parameter with respect to Neptune, which
holds if the planet is on a quasi-circular orbit as in the simulations
of {\it Gomes} (2003). In the Nice model, the eccentricity of Neptune
breaks the conservation of the Tisserand parameter, and hence, in
principle, {inclinations can be larger.}

The results of the simulations based on the Nice model also provide a
qualitative explanation {for} the observed correlations between
inclination and physical properties. The particles that are trapped in
the cold classical belt come, almost exclusively, from the
{outermost parts} of the planetesimal disk { --- in
  particular beyond 29~AU}. Conversely, a significant fraction of
those trapped in the hot population come from the inner disk. Thus, if
one assumes that the largest objects could form only in the inner part
of the disk,{then} these objects can only (or predominantly)
be {found} in the hot population.  Similarly, if one assumes
that (for some unknown reason) the objects from the outer part of the
disk are red and those from the inner part are gray, the cold
population would be composed almost exclusively of red objects,
whereas the hot population would contain a mixture of red and gray
bodies.

The simulations in {\it Levison et al{$.$}} (2007) show that 50
to 130 particles out of 60,000 are trapped in the classical belt (cold
and hot populations together in roughly equal proportion).
{According} to the Nice model, the {original}
planetesimal disk contained 35~$M_\oplus$, {thus this} model
predicts that the classical Kuiper belt should {currently}
contain between $\sim\!0.02$ and $\sim\!0.08 M_\oplus$, in good
agreement from observational estimates.  Of course, to be viable, the
model needs to explain not only the total mass of the belt, but also
the total number of bright, detectable bodies{.  It does this
  quite nicely if one assumes that the original disk
  size-distribution is similar to the one currently observed.  As we
  explained in section~\ref{removal}, this is consistent with other
  constraints like the formation of the Pluto-Charon binary.}  Thus,
the Nice model explains, for the first time, the mass deficit of the
Kuiper belt and the ratio between the hot and the cold population, in
the framework of an initial planetesimal size distribution that
fulfills all the constraints enumerated in section~\ref{depletion}.

\begin{figure}[t!]
\centerline{\psfig{figure=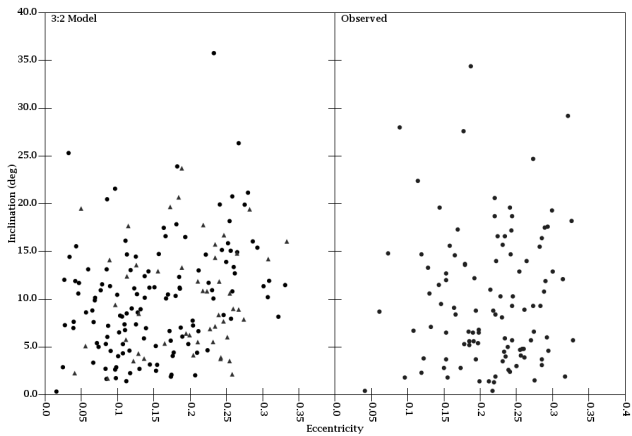,height=5.7cm}}
\vspace*{-.3cm}
\caption{The eccentricity--inclination distribution of the Plutinos. Left
  panel: the simulated distribution; black dots refer to particles from the
  outer disk and gray triangles to particles from the inner disk. Right
  panel: the observed distribution. The relative deficit of observed
Plutinos at low eccentricity with respect to the model is probably due
to observational biases and to the criterion used to select `resonant
objects' from the simulation (see {\it Levison et al.}, 2007).}  \vspace{0.3cm}
\label{2:3}
\end{figure}

Finally, the Nice model reproduces in a satisfactory way the orbital
distributions of the populations in the main mean motion resonances
with Neptune. Fig.~\ref{2:3} compares the ($e,i$) distribution of the
Plutinos obtained in {one of {\it Levison et al$.$ (2007)}'s
  simulations}, against the observed distribution. The overall
agreement is quite {good}. In particular, {this is the
  first model that} does not produce an overabundance of resonant
objects with low inclinations.  

Moreover, the left panel of Fig,~\ref{2:3} {uses different
  symbols to indicate the particles captured from the inner
  ($a<29\,$AU)} and the outer {($a>29\,$AU) parts} of the
disk. As one sees, the particles are very well mixed, which is in
agreement with the absence of correlations between colors and
inclinations among the Plutinos.  Conversely, {a very strong
  correlation was expected in the original {\it Gomes} (2003) scenario
  because a large number of low-inclination bodies were captured} from
the cold disk. The reason {that the Nice model is so much more successful
  than} previous models is that the 2:3 resonance cannot capture any
{objects via} the mechanism of {\it Malhotra} (1995).
{This is due to the fact that} the resonance is already beyond
the disk's outer edge at the beginning of the simulation (i.e. after
the last encounter of Neptune with Uranus).  We believe that this
success {strongly supports that idea} of a planetesimal disk
truncated at 30--35~AU and of a `jump' of Neptune towards the outer
edge of the disk.

As for the higher order resonances beyond 50~AU, the simulations of
{\it Levison et al{$.$}} (2007) produce populations with
moderate libration amplitudes and inclinations consistent with
observations, thus satisfying the constraint posed by {\it Lykawka and
  Mukai} {(2007a).}

{In {summary}, the strength of the Nice model is that it is
able to explain {\it all} the intriguing properties of the Kuiper
belt, at least at a semi-quantitative level, in the framework of a
{single, unique event}. That the same scenario also explains
the orbital architecture of the giant planets, the Trojans {of
both Jupiter and Neptune,} and the LHB is, of course, a non-negligible
additional {plus} that should give {credence} to the
model.}

{Of course, the Nice model is not perfect.  As we have seen above, the
simulations thus far performed have not been able to simultaneously
produce: 1) the very highest inclinations that we see, and 2) a cold
belt that is cold enough. The
numerical simulations contain some simplifications which might effect
the results.  In particular, mutual collisions and collective
gravitational effects among the planetesimals are neglected. Moreover,
as discussed in {\it Levison et al} (2007), only a subset of the
evolutions of the giant planets observed in the simulations of the
{\it Tsiganis et al.} (2005) will produce a cold classical belt.  Some
experiments overly excite inclinations so that a cold belt is not
formed, despite producing good final planetary orbits.  Nonetheless,
we feel that the Nice model's strengths outweigh it weaknesses,
particularly given that other models of Kuiper belt formation have had
much more limited success at reproducing the observations.}


\subsection{Other planetary instability models}

The Nice model is not the first model to make use of a temporary dynamical
instability of the giant planet system (and probably also not the last
one!).

{\it Thommes et al.} (1999) proposed that Uranus and Neptune formed in
between the orbits of Jupiter and Saturn. They were subsequently
destabilized and scattered onto orbits with larger semi-major axis and
eccentricities. The dynamical excitation was eventually damped by the
dynamical friction exerted by a massive planetesimal disk, and the
planets achieved stable orbits. Their simulations showed an
interesting sculpting of the $(a,e)$ distribution in the region
corresponding to the classical Kuiper belt. However, the planetesimal
disk was extended to 60~AU. Thus, no outer edge was produced at the
1:2 resonance with Neptune and there was not enough mass depletion in
the Kuiper belt. Moreover, {since Uranus and Neptune started
  between Jupiter and Saturn they suffered much stronger encounters
  with the gas giants than occurred in the Nice model.  As a result,
  the planetesimal disk needed to be much more massive --- so massive
  that} if the simulations had been run to completion, Neptune would
have migrated well beyond 30~AU.

{\it Chiang et al.} (2006) have recently speculated on a scenario
based on recent work by {\it Goldreich et al$.$} (2004a,b), who, from
analytic considerations, predicted the formation of 5 planets between
20 and 40 AU. These planets remained stable during their formation
because their orbits were continuously damped by the dynamical
friction exerted by a disk of {planetesimals that contained
  more mass than the planetary system.  These planetesimals were very
  small (sub-meter in size) and thus remain dynamically cold due to}
collisional damping. When the planets reached Neptune-mass, the mass
of the planets and the mass of the disk became comparable, so that the
planets became unstable.  {{\it Goldreich et al$.$}}
conjectured that 3 of the 5 planets were ejected and the two remaining
ones stabilized on orbits comparable to those of Uranus and Neptune.

{{\it Chiang et al$.$} suggested that at the time of the
  instability, the disk contained two populations: one made up of
  $\sim\!100\,$km objects and one consisting of sub-meter objects.}
The current Kuiper belt structure would be the result of the orbital
excitation suffered during the multi-planet instability. {The
  hot population would be made up of the larger objects which were
  permanently excited during the instability.  In contrast the
  smallest planetesimals would suffer a significant amount of
  collisional damping, which would have led} to the eventual accretion
of the cold population.  Numerical simulations made by {\it Levison
  and Morbidelli} (2007) with a new code that accounts for a
planetesimal disk with strong internal collisional damping, invalidate
{the {\it Goldreich et al$.$}} proposal. It is found that a
system of 5 unstable Neptune-mass planets systematically leads to a
system with more than 2 planets, spread in semi-major axis well beyond
30~AU.  Thus, the architecture of the solar planetary system is
inconsistent with {{\it Goldreich et al$.$}'s idea.}

\section{Conclusions and Discussion}
\label{conclusions}

In this chapter we have tried to understand which kind of solar system
evolution could have produced the most important properties of the
Kuiper belt: its mass deficit, its outer edge, the co-existence of a
cold and a hot classical population with different physical
properties, {and} the presence of resonant {populations.
  We} have proceeded by basic steps, trying to narrow the
number of possibilities by considering one Kuiper belt feature after
the other, and starting from the most accepted dynamical process
(planet migration) {and eventually ending with} a more
extravagant one (a temporary instability of the giant planets).

We have converged to a basic scenario with three ingredients: the
planetesimal disk was truncated close to 30~AU and the Kuiper belt was
initially empty; the size distribution in the planetesimal disk was
similar to the current one in the Kuiper belt, but the number of
objects at each size was larger by a factor of $\sim 1,000$; the
Kuiper belt objects are just a very small fraction of the original
planetesimal disk population and were implanted onto their current
orbit from the disk during the evolution of the planets. A temporary
high-eccentricity phase of Neptune, when the planet was already at
$\sim 28$~AU --as in the Nice model-- {seems to be the best way} to
implant the cold population.

{The mass deficit problem is the main issue that drove us to this
  conclusion.  In particular, we started from the consideration that,
  if a massive disk} had extended into the Kuiper belt, neither
  collisional grinding nor dynamical ejection could have depleted its
  {mass to current levels}.  Dynamical ejection seems to be excluded
  by the constraint that {Neptune did not migrate past $30\,$AU.}
  Collisional grinding seems to be excluded by the {arguments} that
  the size distribution was about the same everywhere in the disk {and
  that} $\sim\!1,000$ Pluto-size bodies had to {exist} in the
  planetary region.

Is there a flaw in this reasoning? Are we really sure that the cold
  population did not form in situ?  
  The argument that
  the size distribution of the in-situ population should be similar to
  that in the region spanned by Neptune's migration neglects possible
  effects due to the presence of an edge.  After all, an edge is a big
  discontinuity in the size and mass distribution, so that it may not
  be unreasonable that the region adjacent to the edge had very
  different properties from the region further away from the edge.

Our view of the Kuiper belt evolution could radically change if a
model of accretion were developed that produces {a disk of
planetesimals with a size distribution that changes drastically with
distance, such that (a) beyond 45 AU all objects are too small to be
detected by telescope surveys, (b) in the 35--45 AU region the
distribution of the largest objects is similar to that observed in the
current cold population {while} most of the mass {is}
contained in small bodies{,} and (c) within 35~AU most of the
mass is contained in large bodies {and} the size distribution
culminating with $\sim 1,000$ Pluto-sized objects. If this were the
case, the disk beyond 35 AU could lose most of its mass by collisional
grinding before the beginning of Neptune's migration,
{particularly} if the latter was triggered {late as} in
our LHB scenario (see {sect.~3.4} of the {\it chapter} by
Kenyon et al.). Within 35 AU, because of the different size
distribution, collisional grinding would have been ineffective ({\it
Charnoz and Morbidelli}, 2007). Therefore, at the {time of the
LHB}, the system would have been similar to the one required by the
Nice model{, in that} Neptune would have seen an effective edge
in the planetesimal mass distribution that would have {kept it
from migrating beyond} 30~AU.}

{Whether {the spatial variation we described above of} the
size distribution in the planetesimal disk is reasonable or not is
beyond our current understanding. The {\it chapter} by Kenyon et
{al.~in} this book, nicely shows that the coagulation/erosion
process is always on the edge of an instability. In fact, the
dispersion velocity of the small bodies is of the order of the escape
velocity from the largest bodies. Depending on the details of the
collisional cascade (see {sect.~3.5} in that chapter) the
dispersion velocity can be slightly smaller than the escape velocities
(favoring the accretion of a large number of massive bodies and
producing a top-heavy size distribution which does not allow an
effective collisional grinding), or can be slightly larger (stalling
runaway accretion and leaving most of the mass in small, easy to break
bodies). {Perhaps}, the inner part of the disk was in the first
regime and the outer part was in the second one, with a relatively
sharp transition zone between the two parts.  More work is needed to
clarify the situation, with a close collaboration between experts of
accretion and of dynamical evolution.}

\begin{figure}[t!]
\centerline{\psfig{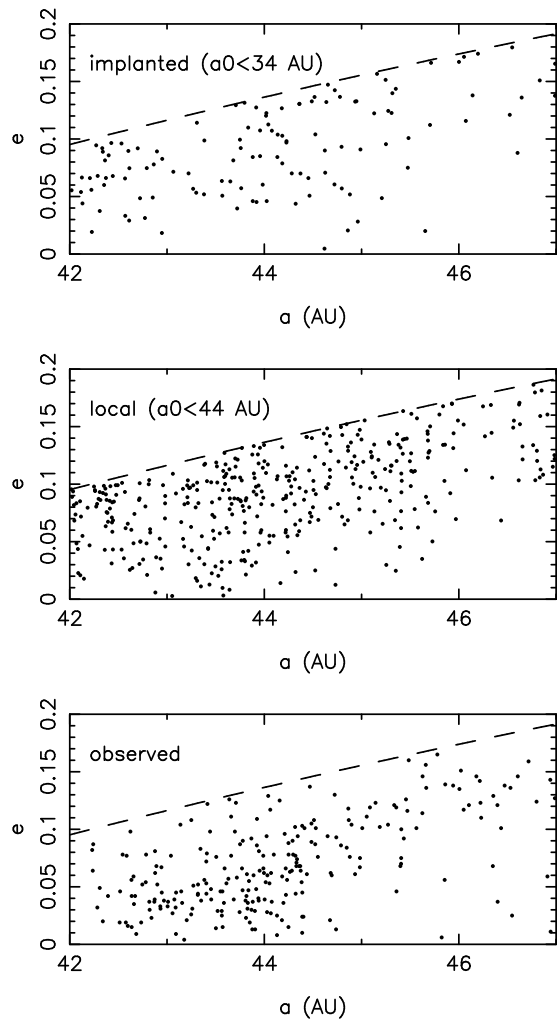}}
\vspace*{-.3cm}
\caption{{The semi-major axis vs. eccentricity distribution in the
  classical belt ($q>38$~AU). Top panel: the distribution of the
  objects {that originated} within 34~AU, implanted in the Kuiper belt
  at the time of the LHB. Middle panel: the distribution of the bodies
  {assuming the disk originally extended} to 44~AU. Bottom
panel: the observed distribution.}}  \vspace{0.3cm} \label{comp_all}
\end{figure}

{From a purely dynamical point of view, the Nice model is not
inconsistent with the existence of a local, low mass Kuiper belt
population, extended up to 44--45~AU. In fact, Fig.~\ref{comp_all}
compares the $(a,e)$ distribution observed in the classical belt with
$q>38$~AU (bottom panel) with the one that the Nice model predicts
assuming that the disk was truncated at 34~AU (top panel, enlargment
of the left panel in Fig.~\ref{Nice-ae}) or assuming that the disk was
truncated at 44~AU (middle panel). The two model distributions are
statistically equivalent, and are both very similar to the observed
distribution.  In the case {where the outer edge was placed} at
44--45~AU about 7\% of the particles initially in the Kuiper belt
($a>40$~AU, $q>38$~AU) remain there, although {their orbits
have been modified}. The {others} escape to larger
eccentricities during the phase when the Kuiper belt is globally
unstable, discussed in sect.~7.  If 90\% of the local mass escapes,
the local belt had to {have been signficantly depleted before}
the time of the LHB, probably accounting for only a few tens of an
Earth mass, otherwise, presumably (see {sect.~6.1}), Neptune
would have been driven past 30~AU. }

{We end this chapter by encouraging observers to attempt to
  probe the regions beyond 50~AU.  In particular, even if} the absence of a
population of objects similar to that in the 40--50~AU region is now
secure, nothing is known on the possible existence of small objects.
Recent claims on a large number of stellar occultation events by
distant $\sim 100$m objects ({\it Roques et al.}, 2006; {\it Chang et
  al.}, 2006; {\it Georgevits}, 2006) may suggest, if confirmed { (see
  {\it Jones et al.} 2006 for a rebuttal of {\it Chang et al.} results)}, the
existence of an extended disk of small planetesimals that {did}
not grow to directly detectable sizes. If true, this would give us
extremely valuable information on the origin of the Kuiper belt edge. 
The model by Weidenschilling predicts the existence of an extended disk
of exclusively small bodies, although of sizes up to $\sim 1 m$ rather than
$\sim 100$~m (see Fig.~\ref{stu}).  Conversely, other
models on edge formation, such as the disk stripping by passing stars,
photo-evaporation or turbulent stirring, would predict the total
absence of objects, of any size. Therefore we encourage the pursuit of
stellar occultation programs, until the real situation is clarified.

So far, the Kuiper belt has taught a lot to us about the evolution of
the planets. Planet migration, for instance, had been totally
overlooked despite the pioneer work by Fernandez and Ip, until the
Kuiper belt was discovered. Moreover, as we mentioned above, we can
find in the Kuiper belt further evidence that the giant planets passed
through a temporary phase of {violent} instability. But, the
Kuiper belt can do more. It can {potentially} teach us about
planetesimal formation and the growth of larger objects
{because the objects that inhabit it most likely probe
  different regions of the proto-planetary disk where accretion
  proceeded in very different ways.}  Thus, it is {a
  dreamed-of} laboratory to test and calibrate the accretion models.
This is probably the main venue for the future.

\acknowledgments

{H.F.L$.$ thanks NASA's {\it Origins} and {\it Planetary Geology and
Geophysics} programs for supporting his involvement in the research
related to this Chapter.  A.M$.$ is also grateful to the French
National Council of Scientific Research (CNRS) and National Programme
of Planetology (PNP) for support.  RSG thanks the Brasilian National
  Council for Science and Technology (CNPq) for support.}

\centerline\textbf{ REFERENCES}
\bigskip
\parskip=0pt
{\small
\baselineskip=11pt

\refs Adams, F.~C., Hollenbach, 
D., Laughlin, G., Gorti, U.\ 2004.\ Photoevaporation of Circumstellar Disks 
Due to External Far-Ultraviolet Radiation in Stellar Aggregates.\ 
Astrophysical Journal 611, 360-379. 

\refs Agnor, C.~B., 
Hamilton, D.~P.\ 2006.\ Neptune's capture of its moon Triton in a 
binary-planet gravitational encounter.\ Nature 441, 192-194. 

\refs Allen, R.L., Bernstein, G.M., Malhotra, R. 2001. The edge of
  the solar system.  {\it Astroph. J.}, {\bf 549}, L241-L244.

\refs Allen, R.L., Bernstein, G.M., Malhotra, R. 2002. Observational Limits
  on a Distant Cold Kuiper Belt. {\it Astron. J.}, {\bf 124}, 2949-2954.

\refs Augereau, J.-C., 
Beust, H.\ 2006.\ On the AU Microscopii debris disk. Density profiles, 
grain properties, and dust dynamics.\ Astronomy and Astrophysics 455, 
987-999. 

\refs Benz, W., Asphaug, 
E.\ 1999.\ Catastrophic Disruptions Revisited.\ Icarus 142, 5-20. 

\refs Barkume, K. Brown, M. and Schaller, E.L. 2006. Discovery of a
colisional family in the Kuiper belt. BAAS, 38, 565.
  
\refs Beauge, C.\ 1994.\ Asymmetric 
liberations in exterior resonances.\ Celestial Mechanics and Dynamical 
Astronomy 60, 225-248. 

\refs Bernstein, G.~M., Trilling, D.~E., Allen, R.~L., Brown,
M.~E., Holman, M., Malhotra, R.\ 2004.\ The Size Distribution of
Trans-Neptunian Bodies.\ Astronomical Journal 128, 1364-1390.

\refs Brown M. 2001.
The Inclination Distribution of the Kuiper Belt.
{\it Astron. J.}, {\bf 121}, 2804-2814.

\refs Brown, M.~E., Trujillo, 
C.~A., Rabinowitz, D.~L.\ 2005.\ Discovery of a Planetary-sized Object in 
the Scattered Kuiper Belt.\ Astrophysical Journal 635, L97-L100. 

\refs Brunini A., Melita M. 2002. The existence of a planet beyond 50 AU
  and the orbital distribution of the classical Edgeworth Kuiper 
  belt objects. {\it Icarus}, {\bf 160}, 32-43.


\refs Canup, R.~M., Ward, 
W.~R.\ 2006.\ A common mass scaling for satellite systems of gaseous 
planets.\ Nature 441, 834-839. 

\refs Charnoz, S. and Morbidelli, A. 2007. Coupling dynamical and
collisional evolution of small bodies. II: Forming the Kuiper Belt,
the scattered disk and the Oort cloud. Icarus in press.

\refs Chiang., E., Lithwick, Y., Murray-Clay, R., Buie, M., Grundy,
W. and Holman, M. 2006. A brief history of the trans-Neptunian
space. In {\it Protostars and planets
  V}, (B. Reipurth et al. eds.), University Arizona Press, Tucson, Az.  

\refs Chirikov, B.~V.\ 1960.\ 
Resonance processes in magnetic traps.\ Journal of Plasma Physics 1, 
253-260. 

\refs Davis D. R., Farinella P. 1998, Collisional Erosion of a 
  Massive Edgeworth-Kuiper Belt:
  Constraints on the Initial Population. In 
  {\it Lunar Planet. Science Conf}. {\bf 29}, 1437--1438.

\refs Davis D. R., Farinella P. 1997. Collisional
  Evolution of Edgeworth-Kuiper Belt Objects. {\it Icarus}, {\bf 125}, 50--60.

\refs Dermott, S.~F., 
Murray, C.~D.\ 1983.\ Nature of the Kirkwood gaps in the asteroid belt.\ 
Nature 301, 201-205. 

\refs Duncan, M. J., Levison, H. F., Budd, S. M. 1995, 
  The long-term stability of
  orbits in the Kuiper belt, {\it Astron. J.}, {\bf 110}, 3073--3083. 

\refs Duncan, M.~J., 
Levison, H.~F.\ 1997.\ A scattered comet disk and the origin of Jupiter 
family comets.\ Science 276, 1670-1672. 

\refs Emel'yanenko, 
V.~V., Asher, D.~J., Bailey, M.~E.\ 2003.\ A new class of trans-Neptunian 
objects in high-eccentricity orbits.\ Monthly Notices of the Royal 
Astronomical Society 338, 443-451. 

\refs Fernandez, J.~A., Ip, 
W.-H.\ 1984.\ Some dynamical aspects of the accretion of Uranus and Neptune 
- The exchange of orbital angular momentum with planetesimals.\ Icarus 58, 
109-120.

\refs Georgevits, G.\ 2006.\ 
Detection of Small Kuiper Belt Objects by Stellar Occultation.\ 
AAS/Division for Planetary Sciences Meeting Abstracts 38, \#37.07. 

\refs Gladman, B., Kavelaars, J.J., Petit, J.M., Morbidelli, A.,
Holman, M.J., Loredo, Y., 2001. The structure of the Kuiper belt: Size
distribution and radial extent. {\it Astron. J.}, {\bf 122},
1051-1066.

\refs Goldreich, P., 
Lithwick, Y., Sari, R.\ 2004a.\ Final Stages of Planet Formation.\ 
Astrophysical Journal 614, 497-507.

\refs Goldreich, P., 
Lithwick, Y., Sari, R.\ 2004b.\ Planet Formation by Coagulation: A Focus on 
Uranus and Neptune.\ Annual Review of Astronomy and Astrophysics 42, 
549-601. 

\refs Gomes R. S. 2000. Planetary Migration and Plutino Orbital Inclinations
  {\it Astron. J.}, {\bf 120}, 2695-2707. 

\refs Gomes, R.~S.\ 2003.\ The origin 
of the Kuiper Belt high-inclination population.\ Icarus 161, 404-418. 

\refs Gomes, R.~S., Morbidelli, 
A., Levison, H.~F.\ 2004.\ Planetary migration in a planetesimal disk: why 
did Neptune stop at 30 AU?.\ Icarus 170, 492-507. 

\refs Gomes, R., Levison, 
H.~F., Tsiganis, K., Morbidelli, A.\ 2005.\ Origin of the cataclysmic Late 
Heavy Bombardment period of the terrestrial planets.\ Nature 435, 466-469. 

\refs Grundy, W.~M., Noll, 
K.~S., Stephens, D.~C.\ 2005.\ Diverse albedos of small trans-neptunian 
objects.\ Icarus 176, 184-191. 

\refs Hahn, J.~M., Malhotra, R.\ 1999.\ Orbital Evolution of Planets
  Embedded in a Planetesimal Disk.\ Astronomical Journal 117, 3041-3053.

\refs Hahn, J.~M., 
Malhotra, R.\ 2005.\ Neptune's Migration into a Stirred-Up Kuiper Belt: A 
Detailed Comparison of Simulations to Observations.\ Astronomical Journal 
130, 2392-2414. 
 
\refs  Henrard J. 1982. Capture into resonance - An
  extension of the use of adiabatic invariants. {\it Cel. Mech.}, {\bf 27},
  3--22.

\refs Hollenbach, D., 
Adams, F.~C.\ 2004.\ Dispersal of Disks Around Young Stars: Constraints on 
Kuiper Belt Formation.\ ASP Conf.~Ser.~324: Debris Disks and the Formation 
of Planets 324, 168. 

\refs Holman, M.~J., 
Wisdom, J.\ 1993.\ Dynamical stability in the outer solar system and the 
delivery of short period comets.\ Astronomical Journal 105, 1987-1999. 

\refs Ida S., Larwood J., Burkert A. 2000.
  Evidence for Early Stellar Encounters in the Orbital 
  Distribution of Edgeworth-Kuiper Belt Objects.
  {\it Astroph. J.}, {\bf 528}, 351--356.

\refs Jones, T.~A., Levine, 
A.~M., Morgan, E.~H., Rappaport, S.\ 2006.\ Millisecond Dips in Sco X-1 are 
Likely the Result of High-Energy Particle Events.\ The Astronomer's 
Telegram 949, 1. 

\refs Kenyon, S.J., Luu, J.X.  1998. Accretion in the early Kuiper belt: I.
Coagulation and velocity evolution. {\it Astron. J.}, {\bf 115}, 2136-2160.

\refs Kenyon, S.J., Luu, J.X.  1999a. Accretion in the early Kuiper belt: II.
Fragmentation. {\it Astron. J.}, {\bf 118}, 1101-1119.

\refs Kenyon, S.J., Luu, J.X.  1999b. Accretion in the early outer solar system.
 {\it Astrophys. J}, {\bf 526}, 465-470

\refs Kenyon S.J. and Bromley B.C. 2002.  Collisional Cascades in
  Planetesimal Disks. I. Stellar Flybys. {\it Astron. J.}, {\bf 2002},
  1757-1775. 

\refs Kenyon, S.~J., 
Bromley, B.~C.\ 2004a.\ The Size Distribution of Kuiper Belt Objects.\ 
Astronomical Journal 128, 1916-1926. 

\refs Kenyon, S.~J., 
Bromley, B.~C.\ 2004b.\ Stellar encounters as the origin of distant Solar 
System objects in highly eccentric orbits.\ Nature 432, 598-602. 

\refs Kne{\v z}evi\'c Z., Milani A., Farinella P.,
Froeschl\'e Ch. and Froeschl\'e C.  1991. Secular resonances from 2
to 50 AU. {\it Icarus}, {\bf 93}, 316-330.

\refs Kobayashi H., Ida S. 2001. The Effects of a Stellar Encounter 
  on a Planetesimal Disk. {\it Icarus}, {\bf 153}, 416-429. 

\refs Kobayashi, H., Ida, 
S., Tanaka, H.\ 2005.\ The evidence of an early stellar encounter in 
Edgeworth Kuiper belt.\ Icarus 177, 246-255. 

\refs Kuchner M.J., Brown M.E., Holman M. 2002.
  Long-Term Dynamics and the Orbital Inclinations of the Classical Kuiper Belt
  Objects. {\it Astron. J.}, {\bf 124}, 1221-1230.

\refs Levison H.F., Stern S.A. 2001. On the Size Dependence of the
  Inclination Distribution of the Main Kuiper Belt. {Astronomical J.}, {\bf
    121}, 1730-1735.

\refs Levison, H.~F., 
Morbidelli, A.\ 2003.\ The formation of the Kuiper belt by the outward 
transport of bodies during Neptune's migration.\ Nature 426, 419-421. 

\refs Levison, H.~F., 
Morbidelli, A., Dones, L.\ 2004.\ Sculpting the Kuiper Belt by a Stellar 
Encounter: Constraints from the Oort Cloud and Scattered Disk.\ 
Astronomical Journal 128, 2553-2563. 

\refs Levison, H.~F., Morbidelli, A., Gomes, R. and Backman, D. 2006. 
Planet migration in planetesimal disks. In {\it Protostars and planets
  V}, (B. Reipurth et al. eds.), University Arizona Press, Tucson, Az.  

\refs Levison, H.~F. and Morbidelli, A. 2007. Neptune-mass planets did
not form in the Kuiper belt. Icarus, in press.

\refs Levison H.~F., Morbidelli, A., Gomes, R. and Tsiganis, K. 2007
Origin of the structure of the Kuiper Belt during {a
    Dynamical Instability in the Orbits of Uranus and
    Neptune}. Icarus, submitted.

\refs Lykawka, P.~S. and Mukai, T. 2007a. Evidence for an excited Kuiper
belt of 50 AU radius in the first Myr of solar system history. Icarus,
in press.

\refs Lykawka, P.~S. and Mukai, T. 2007b. Dynamical classification
of trans-Neptunian objects: Probing their origin, evolution and
interrelation. Icarus, in press.

\refs Malhotra, R.\ 1993.\ The 
Origin of Pluto's Peculiar Orbit.\ Nature 365, 819. 

\refs  Malhotra, R.\ 1995.\ The 
Origin of Pluto's Orbit: Implications for the Solar System Beyond Neptune.\ 
Astronomical Journal 110, 420. 

\refs Malhotra, R.\ 1996.\ The 
Phase Space Structure Near Neptune Resonances in the Kuiper Belt.\ 
Astronomical Journal 111, 504. 

\refs Melita M., Larwood J., Collander-Brown S, Fitzsimmons A.,
  Williams I.P., Brunini A. 2002. The edge of the Edgeworth-Kuiper belt:
  stellar encounter, trans-Plutonian planet or outer limit of the primordial
  solar nebula? In {\it Asteroid, Comet, Meteors}, ESA Spec. Publ. series,
  305-308. 

\refs Message, P.~J.\ 1958.\ 
Proceedings of the Celestial Mechanics Conference: The search for 
asymmetric periodic orbits in the restricted problem of three bodies.\ 
Astronomical Journal 63, 443. 

\refs Morbidelli, A., 
Henrard, J.\ 1991.\ The main secular resonances nu6, nu5 and nu16 in the 
asteroid belt.\ Celestial Mechanics and Dynamical Astronomy 51, 169-197. 

\refs Morbidelli A., Thomas F. and Moons M.  1995. The
resonant structure  of the Kuiper belt and the dynamics of the first
five trans-Neptunian objects. {\it Icarus}, {\bf 118}, 322. 

\refs Morbidelli A., Valsecchi G. B. 1997. Neptune
  scattered planetesimals could have sculpted the primordial
  Edgeworth--Kuiper belt, {\it Icarus}, {\bf 128}, 464--468.

\refs Morbidelli A., 2002. {\it Modern Celestial Mechanics:
aspects of Solar System dynamics}, in ``Advances in Astronomy and
Astrophysics'', Taylor \& Francis, London.

\refs Morbidelli, A., 
Levison, H.~F.\ 2004.\ Scenarios for the Origin of the Orbits of the 
Trans-Neptunian Objects 2000 CR$_{105}$ and 2003 VB$_{12}$ (Sedna).\ 
Astronomical Journal 128, 2564-2576. 

\refs Morbidelli, A., 
Levison, H.~F., Tsiganis, K., Gomes, R.\ 2005.\ Chaotic capture of 
Jupiter's Trojan asteroids in the early Solar System.\ Nature 435, 462-465. 

\refs Murray-Clay, R.~A., Chiang, E.~I.\ {2006.\ Brownian
  Motion in Planetary Migration.\ Astrophysical Journal 651,
  1194-1208. }

\refs Nagasawa M., Ida S. 2000.
  Sweeping Secular Resonances in the Kuiper Belt Caused by Depletion of 
  the Solar Nebula.
  {\it Astron. J.}, {\bf 120}, 3311--3322.

\refs Nesvorn\'y D. and Roig F.  2000. Mean motion resonances 
in the trans-neptunian region: Part I: The
2:3 resonance with Neptune. {\it Icarus}, 148, 282-300. 

\refs Nesvorn\'y D. and Roig F.  2001. 
Mean motion resonances in the trans-neptunian region: Part II: the 1:2, 
3:4 and weaker resonances. {\it Icarus}, 150, 104-123. 

\refs Petit, J.-M., Morbidelli, 
A., Chambers, J.\ 2001.\ The Primordial Excitation and Clearing of the 
Asteroid Belt.\ Icarus 153, 338-347. 

\refs Roques, F., and 17 
colleagues 2006.\ Exploration of the Kuiper Belt by High-Precision 
Photometric Stellar Occultations: First Results.\ Astronomical Journal 132, 
819-822. 

\refs Ruden, S.~P., 
Pollack, J.~B.\ 1991.\ The dynamical evolution of the protosolar nebula.\ 
Astrophysical Journal 375, 740-760. 

\refs Sekiya, M.\ 1983.\ 
Gravitational instabilities in a dust-gas layer and formation of 
planetesimals in the solar nebula.\ Progress of Theoretical Physics 69, 
1116-1130.

\refs  Sheppard, S.~S., 
Trujillo, C.~A.\ 2006.\ A Thick Cloud of Neptune Trojans and Their Colors.\ 
Science 313, 511-514. 

\refs Stern, S. A. 1991, On the number of planets in the outer
  solar system - Evidence of a substantial population of 1000-km bodies,
  Icarus, 90, 271--281

\refs Stern, S. A. 1996, On the Collisional Environment, 
  Accretion Time Scales, and Architecture of the
  Massive, Primordial Kuiper
  Belt., {\it Astron. J.}, {\bf 112}, 1203--1210.

\refs Stern, S. A., Colwell, J. E.  1997a,
 Accretion in the Edgeworth-Kuiper Belt: Forming 100-1000 KM Radius Bodies at
 30 AU and Beyond. {\it Astron. J.}, {\bf 114}, 841-849.
 
\refs Stern, S. A., Colwell, J. E.  1997b, Collisional
  Erosion in the Primordial Edgeworth-Kuiper Belt and the Generation of the
  30-50 AU Kuiper Gap, {\it Astroph. J.}, {\bf 490}, 879--885.

\refs Stern, S.~A., 
Weissman, P.~R.\ 2001.\ Rapid collisional evolution of comets during the 
formation of the Oort cloud.\ Nature 409, 589-591.

\refs Stone, J. M.,
Gammie, C. F., Balbus, S. A. and Hawley, J. F. in Protostars and
Planets IV (eds Mannings, V., Boss, A. P. and Russell, S. S.) 589
(Univ. Arizona Press, Tucson, 1998)

\refs Thommes, E.~W., Duncan, 
M.~J., Levison, H.~F.\ 1999.\ The formation of Uranus and Neptune in the 
Jupiter-Saturn region of the Solar System.\ Nature 402, 635-638.

\refs Trujillo C.A., Brown M.E. 2001.  The Radial Distribution of 
the Kuiper Belt. {\it Astroph. J}, {\bf 554}, 95-98. 

\refs Trujillo, C.~A., 
Brown, M.~E.\ 2003.\ The Caltech Wide Area Sky Survey.\ Earth Moon and 
Planets 92, 99-112. 

\refs Tsiganis, K., Gomes, 
R., Morbidelli, A., Levison, H.~F.\ 2005.\ Origin of the orbital 
architecture of the giant planets of the Solar System.\ Nature 435, 
459-461. 
 
\refs Youdin, A.~N., Shu, 
F.~H.\ 2002.\ Planetesimal Formation by Gravitational Instability.\ 
Astrophysical Journal 580, 494-505. 

\refs Weidenschilling S. 2003. Formation of
  Planetesimals/Cometesimals in the Solar nebula. in {\it Comet II}, Festou et
  al. eds., University Arizona Press, Tucson, Az. 97--104.

\refs Williams, J.~G., 
Faulkner, J.\ 1981.\ The positions of secular resonance surfaces.\ Icarus 
46, 390-399. 

\refs Wisdom, J.\ 1980.\ The 
resonance overlap criterion and the onset of stochastic behavior in the 
restricted three-body problem.\ Astronomical Journal 85, 1122-1133. 

\end{document}